\documentclass[%
 reprint,
 amsmath,amssymb,
prb,
showkeys
]{revtex4-2}
\UseRawInputEncoding
\usepackage{multirow}
\usepackage{graphicx}
\usepackage{dcolumn}
\usepackage{bm}
\usepackage{longtable}
\usepackage{geometry}
\usepackage{soul}
\usepackage{xcolor}
\usepackage{lineno}
\begin{document}

\preprint{APS/123-QED}

\title{Classification and design of two-dimensional altermagnets}

\author{Sike Zeng$^{1}$}
\author{Dong Liu$^{1}$}
\author{Hongjie Peng$^{1}$}

\author{Chang-Chun He$^{1,2}$}

\author{Xiao-Bao Yang$^{1}$}
\author{Yu-Jun Zhao$^{1}$}%
\email{zhaoyj@scut.edu.cn}

\affiliation{$^{1}$Department of Physics, South China University of Technology, Guangzhou 510640, China \\ $^{2}$Guangdong Provincial Key Laboratory of Functional and Intelligent Hybrid Materials and Devices, South China University of Technology,  Guangzhou 510640, China}
\date{\today}
\begin{abstract}
Altermagnets---newly identified collinear antiferromagnets---carry zero net moment with non-relativistic, spin-polarized bands, distilling the best of ferromagnets and antiferromagnets into a single spintronic platform. Shrunking to the two-dimensional limit, they inherit the tunability of two-dimensional crystals while adding symmetry-protected spin splitting, a combination now driving intense experimental interest. Here, we review the symmetry classification of two-dimensional altermagnets based on spin-group theory and survey the growing list of candidate materials, emphasizing those with large spin splitting for experimental realization. We then examine strategies for engineering two-dimensional altermagnetism. This Review aims to consolidate theoretically proposed candidate materials and realization strategies for two-dimensional altermagnets, providing insights for future experimental efforts in this emerging field.

\end{abstract}

\maketitle


\section{\label{sec:level1}introduction}
Magnetism is a time-honored yet continually thriving field that constitutes one of the core research areas in condensed matter physics and underpins a broad range of modern technologies. Most traditional studies of magnetism have focused on ferromagnets and antiferromagnets. Recently, a new type of collinear antiferromagnet known as altermagnet has been proposed\cite{PhysRevX.12.031042}, sparking growing interest in both theoretical and experimental research\cite{PhysRevX.12.040501,adfm.202409327,song2025altermagnets,fender2025altermagnetism,tamangAltermagnetismAltermagnetsBrief2025}. Altermagnets are characterized by vanishing net magnetization, as in antiferromagnets, and the non-relativistic spin-polarized band structure, as in ferromagnets. They are therefore also referred to as spin-splitting antiferromagnets. This phenomenon was initially predicted by several research groups\cite{vsmejkal2020crystal,PhysRevB.75.115103,hayami2019momentum,PhysRevB.102.014422,mazin2021prediction,ma2021multifunctional}. Due to its non-relativistic origin, altermagnetism is governed by spin-group symmetry\cite{PhysRevX.12.031042,PhysRevX.12.021016}. From a symmetry perspective, the opposite-spin sublattices in altermagnets are connected by proper or improper rotations, rather than by spatial inversion or pure translations, as is the case in conventional antiferromagnets. As a result, conventional antiferromagnets exhibit spin-degenerate band structures due to the protection of $\mathcal{PT}$ or $\mathcal{T\tau}$ symmetry, where $\mathcal{P}$ denotes spatial inversion, $\mathcal{T}$ denotes time-reversal, and $\mathcal{\tau}$ is a fractional translation. In contrast, the absence of these symmetries in altermagnets allows for spin splitting in the band structure, similar to ferromagnets.

From an application perspective, altermagnets hold great promise as efficient platforms for next-generation spintronic devices. Conventional spintronics has been primarily based on ferromagnets, owing to their broken time-reversal symmetry and the associated physical responses. However, the presence of stray fields in ferromagnets poses significant constraints on both storage density and scalability. In addition, the presence of a finite net magnetization renders them sensitive to perturbations from external magnetic fields. In contrast, antiferromagnetic materials offer distinctive advantages, including robustness against external magnetic field perturbations, the absence of stray fields, ultrafast spin dynamics. Consequently, antiferromagnetic spintronics, which replaces ferromagnets with antiferromagnets as the spin-dependent components, has attracted considerable attention and emerged as a highly promising field\cite{RevModPhys.90.015005,jungwirth2016antiferromagnetic}. Nevertheless, the absence of strong spin currents and time-reversal symmetry-breaking responses has limited the practical applications of conventional antiferromagnets. As time-reversal symmetry-breaking antiferromagnets, altermagnets integrate the key merits of both ferromagnets and antiferromagnets, offering new opportunities for advancing spintronics.

\begin{figure*}
	
	\centering
	\includegraphics[width=\linewidth]{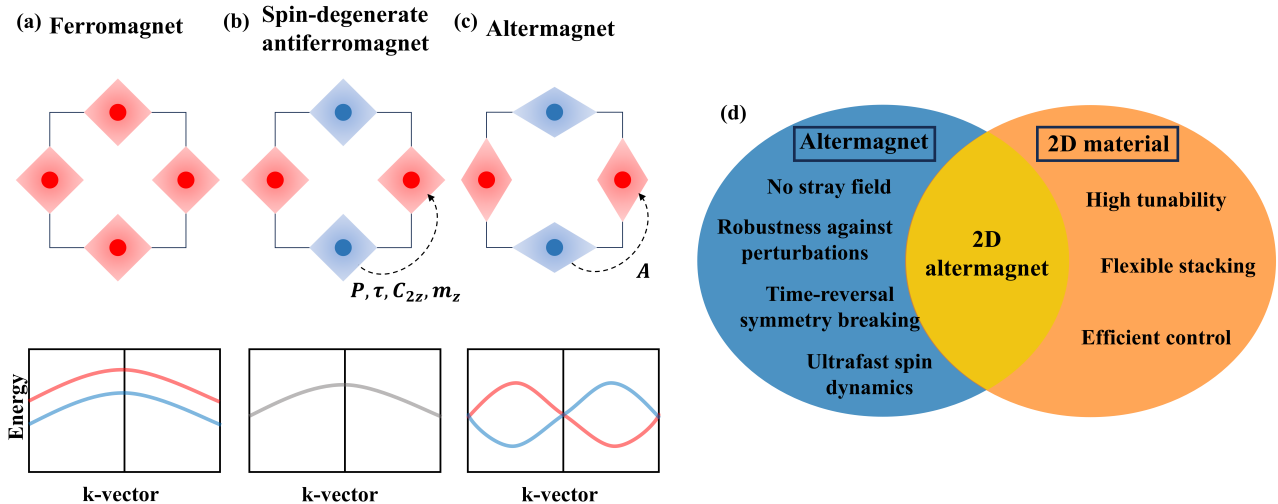}
	\caption{\label{fig:figure1} Schematic illustration of the crystal structures and nonrelativistic band structures of three types of two-dimensional (2D) collinear magnets, and the advantages of two-dimensional altermagnets. (a) Ferromagnets possess a nonzero net magnetization and exhibit spin-split band structures. (b) and (c) Antiferromagnets consist of two magnetic sublattices with opposite spin orientations, resulting in a zero net magnetization. Depending on whether spin splitting appears in the nonrelativistic band structure, they can be classified into spin-degenerate antiferromagnets (i.e., conventional antiferromagnets) and spin-splitting antiferromagnets (i.e., altermagnets). (b) In spin-degenerate antiferromagnets, sublattices are connected by symmetries such as inversion ($\mathcal{P}$), translation ($\tau$), twofold rotation perpendicular to the material plane ($C_{2z}$), or mirror reflection parallel to the material plane ($m_z$), which enforce spin-degenerate bands. (c) In altermagnets the sublattices are related by other proper or improper rotations $A$, leading to spin-split band structures. (d) Schematic diagram illustrating the advantages of 2D altermagnets. The left side highlights the key merits of altermagnets, while the right side showcases the unique advantages of 2D materials. 2D altermagnets retain the key merits of bulk altermagnets, but in contrast to three-dimensional altermagnets, they also leverage the unique advantages of 2D materials.}
	
\end{figure*}

A wide range of materials have been recognized as altermagnets\cite{adfm.202409327}, providing a solid foundation for further theoretical and experimental studies. Through angle-resolved photoemission spectroscopy (ARPES), several materials have been experimentally confirmed as altermagnets, including MnTe\cite{PhysRevLett.132.036702,PhysRevB.109.115102}, CrSb\cite{PhysRevLett.133.206401,zeng2024observation,reimers2024direct,yang2025three}, KV$_2$Se$_2$O\cite{jiang2025metallic}, and Rb$_{1-\delta}$V$_2$Te$_2$O\cite{zhang2025crystal}. Among the experimentally established altermagnets, MnTe stands out as a particularly promising system. Recent nanoscale x-ray magnetic circular dichroism (XMCD) imaging on single-crystal MnTe has provided direct evidence of bulk altermagnetic order, confirming its intrinsic nature beyond thin-film or surface effects\cite{yamamotoAltermagneticNanotexturesRevealed2025}. Spin-dependent transport phenomena, such as the giant magnetoresistance (GMR) effect and the tunnel magnetoresistance (TMR) effect, have been investigated in altermagnets\cite{PhysRevX.12.011028,PhysRevLett.133.056701,PhysRevB.108.L180401,PhysRevLett.126.127701,PhysRevLett.129.137201,PhysRevLett.128.197202}, paving the way for their potential applications in spintronics. In addition, altermagnets have attracted growing research interest in diverse areas, including topological physics\cite{PhysRevB.110.064426,PhysRevB.109.L201109}, ultrafast dynamics\cite{weber2024ultrafast}, valleytronics\cite{PhysRevB.111.094411,PhysRevB.110.L220402,PhysRevB.110.184408}, superconductivity\cite{PhysRevB.109.224502,PhysRevB.110.024503,PhysRevB.108.L060508,PhysRevB.108.224421,cv8s-tk4c,PhysRevB.109.134511}, thermoelectricity\cite{PhysRevB.110.094508,badura2024observation,PhysRevApplied.23.044066,PhysRevB.111.035423,PhysRevB.108.L180401,PhysRevB.110.094427,PhysRevB.111.L020412} and multiferroic\cite{PhysRevLett.134.106801,sun2025proposing,vsmejkal2024altermagnetic,PhysRevLett.134.106802}. 

Following the initial experimental demonstration of two-dimensional (2D) van der Waals (vdW) magnets in 2017\cite{gong2017discovery,huang2017layer}, 2D magnets have attracted widespread attention in both fundamental magnetism and spintronic applications\cite{wang2022magnetic}. Compared to three-dimensional (3D) magnets, they exhibit unique advantages\cite{mak2019probing}, as summarized in the right side of Fig. \ref{fig:figure1}(d). Owing to the weak interlayer vdW interaction, 2D magnets can be combined with any other 2D materials to form heterostructures without suffering from lattice mismatch issues, thereby enabling a wide range of tunable physical properties\cite{huang2020emergent,zhang2025recent}. Even bilayers or multilayers composed of the same material can exhibit emergent phenomena\cite{xu2022coexisting,xie2022twist}. Moreover, 2D magnets exhibit enhanced tunability under external stimuli, such as gating or strain. Furthermore, 2D magnetic materials offer significant promise for enhancing the efficiency of electrical control in magnetic devices\cite{mak2019probing}. 

Building on the concept of two-dimensional magnetic materials, the idea of two-dimensional altermagnets has emerged naturally. 2D collinear magnets can be categorized into three distinct types, with their differences in crystal structures and band structures illustrated in Figs. \ref{fig:figure1}(a)-(c). 2D altermagnets are anticipated to retain the key merits of bulk altermagnets while offering additional advantages that emerge uniquely in reduced dimensions, making them a compelling platform for spintronic applications and for fundamental studies of magnetism, as illustrated in Fig. \ref{fig:figure1}(d). Recently, research on two-dimensional altermagnets has been rapidly growing. The first reported two-dimensional altermagnet is V$_2$Se$_2$O\cite{ma2021multifunctional}, which exhibits a giant noncollinear spin current. As a class of two-dimensional materials, 2D altermagnets have emerged as an important platform for valleytronics research\cite{PhysRevB.110.014442,wu2024valley,li2025ferrovalley}, including studies on valley polarization\cite{PhysRevB.110.L220402,PhysRevB.110.184408,PhysRevB.111.134429,k4x8-84g8} and the valley Hall effect\cite{PhysRevB.111.094411}. Anomalous transport phenomena have also been theoretically predicted in 2D altermagnets\cite{PhysRevB.111.184407,PhysRevB.111.184437,PhysRevB.110.155125}. Moreover, circularly polarized light has been proposed as an effective means to selectively tune the Ruderman-Kittel-Kasuya-Yosida (RKKY) interaction in 2D \emph{d}-wave altermagnets\cite{yarmohammadiAnisotropicLighttailoredRKKY2025}. By combining ferroelectricity, two-dimensional multiferroic altermagnets have been explored\cite{zhu2025two,PhysRevB.110.224418}, where sliding ferroelectricity serves as an efficient method to control the spin splitting in altermagnets\cite{sun2024altermagnetism,sun2025proposing,zhu2025sliding,wang2025two}.The interplay between altermagnetism and superconductivity in a two-dimensional system has been  explored\cite{PhysRevB.108.224421,PhysRevB.108.184505}. It has been shown that when a superconductor is proximitized to an altermagnet, the altermagnetic anisotropic spin splitting can induce a gapless superconducting state\cite{weiGaplessSuperconductingState2024,weiUnconventionalHallEffect2025}. Yet, the field is still dominated by theory, with experimental work lagging behind. Although the 2D limit of altermagnet CdAlSi has been investigated\cite{parfenov2025pushing}, this film is not an altermagnet in principle. Motivated by this gap, we survey all theoretically predicted 2D candidates and the routes that may lead to their laboratory realization, offering experimenters a concise roadmap. 

Broader discussions of altermagnetism can be found in the dedicated reviews cited herein\cite{PhysRevX.12.040501,adfm.202409327,song2025altermagnets,fender2025altermagnetism,tamangAltermagnetismAltermagnetsBrief2025}, which survey the conceptual foundations of altermagnetism, summarize recent experimental and theoretical advances, and highlight the key physical properties, material realizations, and experimental probes associated with this emerging magnetism. Nonetheless, their discussions are centered predominantly on bulk systems. In the 2D limit, enhanced sensitivity to external perturbations, the presence of stacking degrees of freedom, and extensive chemical tunability create design rules and material realizations fundamentally distinct from those in bulk systems, yet these aspects have not been comprehensively consolidated. In this review, we address this gap by introducing a unified symmetry classification for 2D altermagnets based on spin-group theory, compiling all reported monolayer candidates, and summarizing the chemical and structural design strategies that are used in the 2D regime.

\begin{figure}
	
	\centering
	\includegraphics[width=\linewidth]{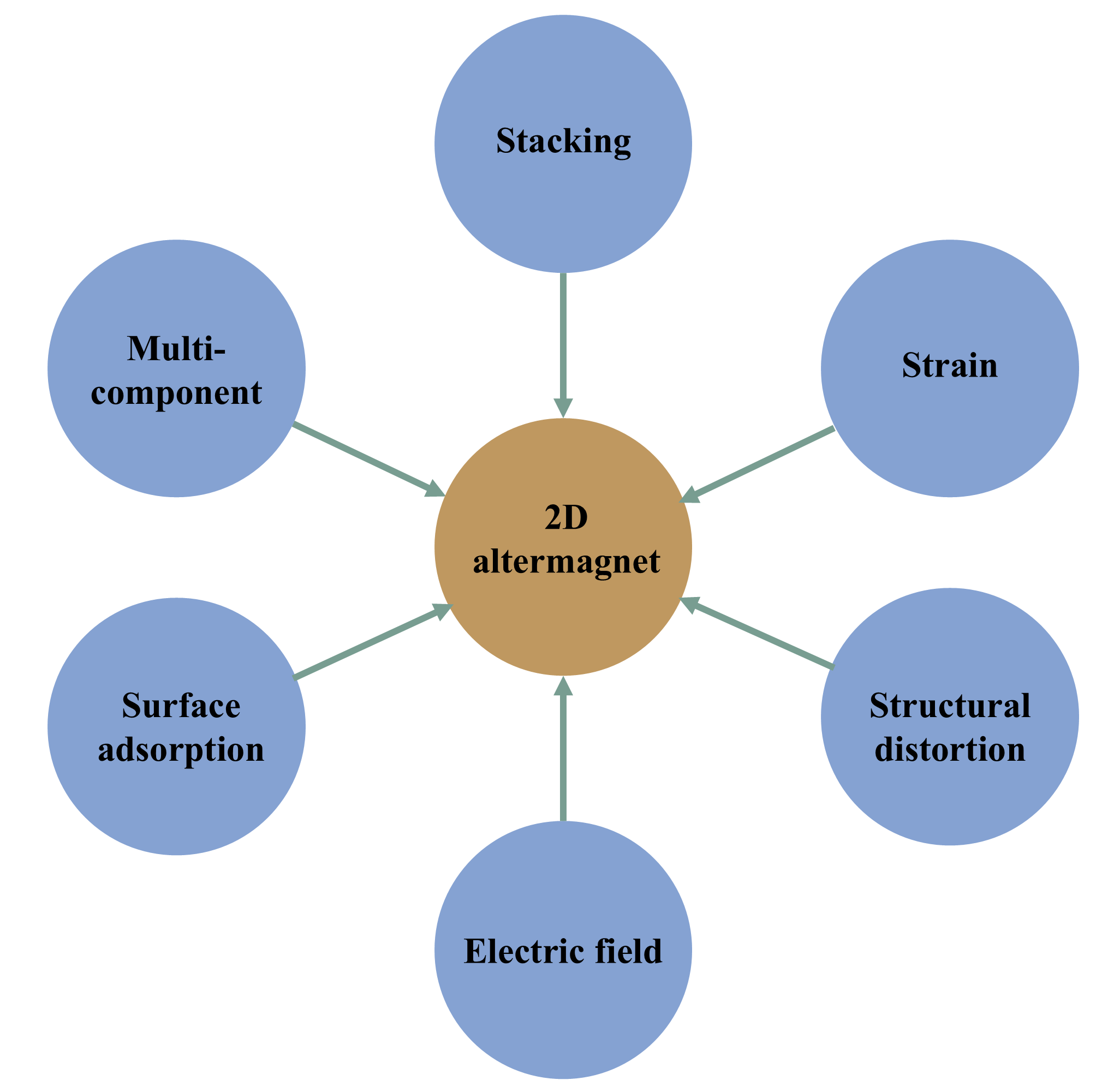}
	\caption{\label{fig:figure01} Schematic illustration of the design routes for two-dimensional (2D) altermagnets. We will introduce six strategies proposed for engineering 2D altermagnetism, including stacking (Section \ref{sec:level4}), multicomponent design (Section \ref{sec:level5}), surface adsorption (Section \ref{sec:level6}), electric-field control (Section \ref{sec:level7}), structural distortion (Section \ref{sec:level8}), and strain (Section \ref{sec:level9}). } 
	
\end{figure}

The Review is organized as follows. We begin by outlining the symmetry framework based on spin groups, which underpins the classification of 2D altermagnets. We then provide an overview of reported 2D altermagnets, summarized in Table \ref{tab:table2}, and highlight representative materials with large spin splitting as promising candidates for experimental realization. In addition, we discuss various strategies proposed for engineering 2D altermagnetism, including stacking, multicomponent design, surface adsorption, electric-field control, structural distortion and strain. A schematic diagram is provided for quick reference, as shown in Fig. \ref{fig:figure01}. Finally, we provide an outlook on the future opportunities and challenges in this emerging field.

\section{\label{sec:level2}Symmetry Classification of two-dimensional altermagnets}

Altermagnetism is a nonrelativistic phenomenon and is therefore governed primarily by spin group symmetries. Spin group theory, originally proposed in the last century\cite{brinkman1966theory,litvin1974spin,litvin1977spin}, has recently been further developed as a powerful framework for describing the symmetries of magnetic structures\cite{PhysRevX.14.031037,PhysRevX.14.031038,PhysRevX.14.031039}. Compared to magnetic groups, spin groups allow for a partial decoupling of real-space and spin-space symmetry operations. In this section, we introduce the symmetry description of two-dimensional altermagnets based on spin group theory.

A spin group is conventionally expressed as the direct product $r_s \otimes R_s$, where $r_s$ denotes the spin-only group consisting of symmetry operations acting solely in spin space, and $R_s$ represents the nontrivial spin group composed of pairs of operations $[R_i \,\|\, R_j]$, in which the operation to the left of the double bar acts exclusively on spin space, while the one on the right acts only on real space. For collinear spin arrangements, the spin-only group is the same and primarily includes two types of symmetry operations: (i) continuous rotations around the common spin axis ($C_{\infty}$); and (ii) a twofold rotation around an axis perpendicular to the spin direction, combined with the inversion symmetry in spin space, which is always realized through time-reversal symmetry, denoted as $[\overline{C_2}\,\|\,\mathcal{T}]$. The second symmetry operation strictly constrains collinear magnetic systems to exhibit only even-parity wave magnetism.

Different types of collinear magnetic orders are characterized by distinct nontrivial spin groups. As three-dimensional groups with two-dimensional translations, layer groups offer a complete symmetry framework for describing the crystal symmetries of quasi-two-dimensional materials. Therefore, for two-dimensional magnetic materials, the real-space part of the nontrivial spin group should be a layer group\cite{PhysRevB.110.054406}. In ferromagnetic systems, only a single spin sublattice is present, as shown in Fig. \ref{fig:figure1}(a). As a result, the corresponding nontrivial spin group takes the form $[E\,\|\,G]$, where $G$ denotes layer group. Although the ferrimagnets contain multiple spin sublattices, there are no symmetry operations connecting them. As a result, their spin groups are identical to those of ferromagnets. Recently, two-dimensional fully compensated ferrimagnets have also attracted increasing attention\cite{PhysRevLett.134.116703}. These ferrimagnets exhibit zero net magnetic moment without symmetry protection, along with nonrelativistic spin-split band structures. Compared to altermagnets, half-metallicity can emerge in the fully compensated ferrimagnets\cite{semboshi2022new}. 

For collinear antiferromagnets, two spin sublattices are present and are connected by a spin-group symmetry operation, which guarantees that the net magnetization vanishes. Therefore, the corresponding nontrivial spin group takes the form \([E\,\|\,H] + [C_2\,\|\,G - H] = [E\,\|\,H] + [C_2\,\|\,AH]\), where H is a halving subgroup of the layer group G, and A is a coset representative. In this case, the operations in the subgroup H connect sites with the same spin orientation, while those in the coset AH connect sites with opposite spin orientations. In two-dimensional systems, spin-degenerate band structures can be protected by four spin-group operations, including $[C_2\,\|\,\mathcal{P}]$, $[C_2\,\|\,\tau]$, $[C_2\,\|\,m_z]$ and $[C_2\,\|\,C_{2z}]$, which is more than in three-dimensional systems\cite{PhysRevB.110.054406}. Here, $\mathcal{P}$ denotes spatial inversion, $\tau$ represents a translation, $m_z$ is the mirror reflection with respect to the plane of the material, $C_{2z}$ denotes a two-fold rotation around the out-of-plane axis. 

\begin{figure}
	
	\centering
	\includegraphics[width=\linewidth]{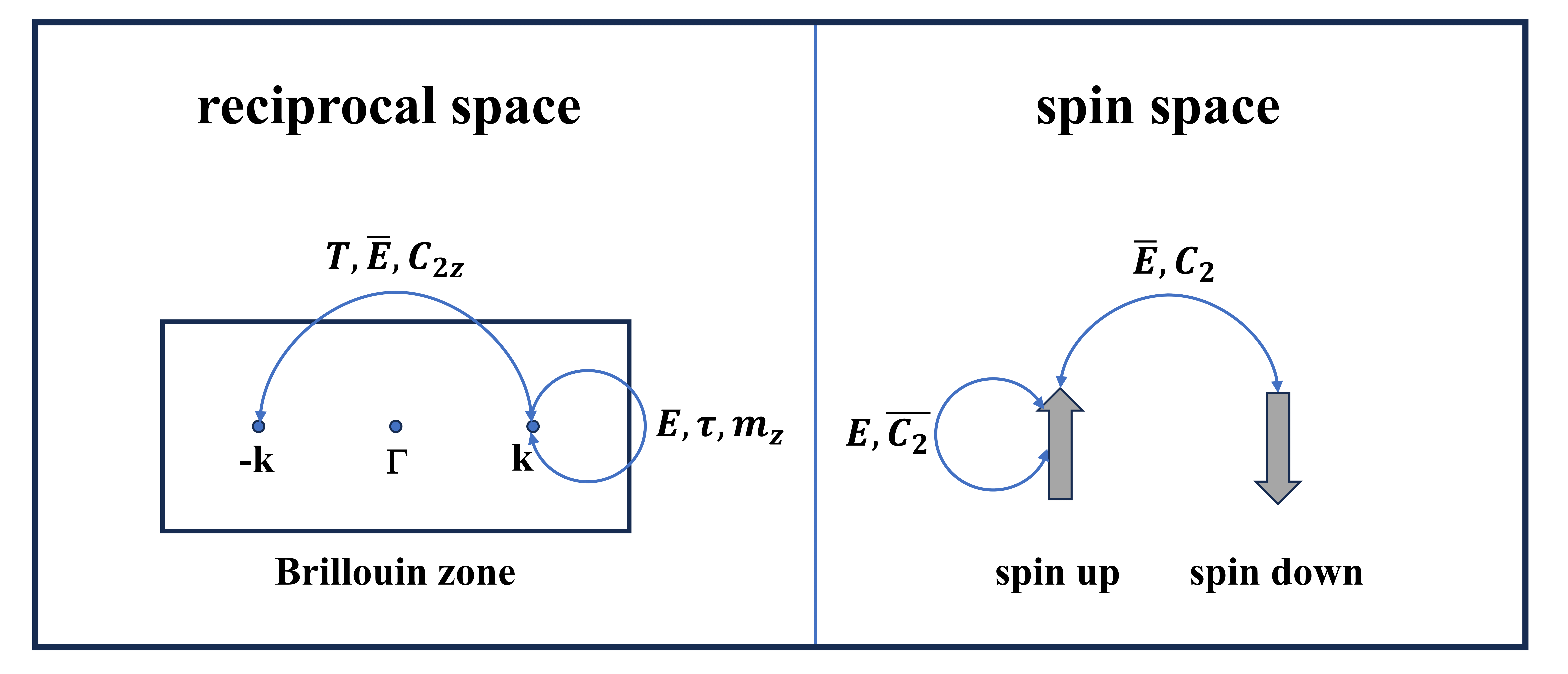}
	\caption{\label{fig:figure0} Schematics of how different spin and spatial symmetries act on the $\varepsilon(s,\mathbf{k})$ spectrum\cite{PhysRevB.110.054406}. In reciprocal space, time reversal ($T$), inversion ($\overline{E}$), and twofold rotation about $z$ ($C_{2z}$) reverse $\mathbf{k}$, whereas identity ($E$), translation ($\tau$), and mirror symmetry about the $xy$ plane ($m_z$) leave $\mathbf{k}$ unchanged. In spin space, inversion ($\overline{E}$) and twofold rotations about axes perpendicular to the spins flip the spin ($C_2$), while identity ($E$) and the combined operation of a perpendicular twofold rotation with spin-space inversion ($\overline{C_2}$) keep it invariant.}
	
\end{figure}

The origin of the spin degeneracy is that each of these operations pairs the spin-up and spin-down states at the same $\mathbf{k}$ point\cite{PhysRevB.110.054406}. Fig. \ref{fig:figure0} provides a schematic illustration to aid the understanding of how various spin-group symmetries affect the $\varepsilon(s,\mathbf{k})$ spectrum. First, $[C_{2}\,\|\,\tau]$ maps the two magnetic sublattices onto each other through a translation $\tau$, and since spin degeneracy does not depend on real-space translations, $[C_{2}\,\|\,\tau]$ is equivalent to $[C_{2}\,\|\,E]$. This directly enforces  
\begin{align}
	\label{eq:1}
[C_{2}\,\|\,\tau]\,\varepsilon(s,\mathbf{k})=\varepsilon(-s,\mathbf{k})=\varepsilon(s,\mathbf{k}).
\end{align}
The second equality holds because this operation is a symmetry of the system. Moreover, $[C_{2}\,\|\,\mathcal{P}]$ combines with the spin-only operation $[\overline{C}_{2}\,\|\,\mathcal{T}]$ to produce  
\begin{align}
	\label{eq:2}
[C_{2}\,\|\,\mathcal{P}]\,[\overline{C}_{2}\,\|\,T]=[\overline{E}\,\|\,\mathcal{PT}],
\end{align}
which corresponds to $\mathcal{PT}$ symmetry that guarantees Kramers degeneracy.
In two dimensions, two additional operations---$[C_{2}\,\|\,m_{z}]$ and $[C_{2}\,\|\,C_{2z}]$---protect the spin degeneracy. This is because $[C_{2}\,\|\,m_{z}]$ leads to 
\begin{align}
	\label{eq:3}
[C_2||m_z]\varepsilon(s,\textbf{k})=\varepsilon(-s,\textbf{k})=\varepsilon(s,\mathbf{k}),
\end{align}
and $[C_{2}\,\|\,C_{2z}]$ combining with $[\overline{C}_{2}\,\|\,\mathcal{T}]$ leads to 
\begin{align}
	\label{eq:4}
[\overline{E}||TC_{2z}]\varepsilon(s,\textbf{k})=\varepsilon(-s,\textbf{k})=\varepsilon(s,\textbf{k}).
\end{align}
Taken together, these symmetry relations reveal that reduced dimensionality naturally expands the set of operations able to protect spin degeneracy in 2D materials.

Therefore, when A belongs to \{$\mathcal{P},\tau,m_z,C_{2z}$\}, the nonrelativistic band structure of the antiferromagnet is spin-degenerate, as shown in Fig. \ref{fig:figure1}(b). In other words, two dimensional collinear antiferromagnets in which opposite-spin sites are connected by one of the operations in \{$\mathcal{P},\tau,m_z,C_{2z}$\} exhibit spin-degenerate band structures in the nonrelativistic limit. Recently, antiferromagnets in which opposite-spin sites are connected by $m_z$ or $C_{2z}$ have been classified as type-IV two-dimensional collinear magnets\cite{rn1l-d6cq}. These systems are characterized by spin-degenerate band structures in the nonrelativistic limit, and simultaneously exhibiting time-reversal symmetry-breaking responses. 

Since the symmetry operations of the form $[C_2\,\|\,A]$, with $A \in \{\mathcal{P}, \tau, m_z, C_{2z}\}$, enforce spin degeneracy, they are incompatible with the nonrelativistic spin splitting that characterizes altermagnets, as shown in Fig. \ref{fig:figure1}(c). Therefore, the spin groups that describe 2D altermagnets must exclude such symmetry operations. Since our classification of 2D altermagnets focuses on spin-momentum locking, translational operations can be neglected. For all collinear magnets, regardless of whether they possess real-space inversion symmetry, the nonrelativistic band structure remains invariant under real-space inversion operations\cite{PhysRevB.110.054406}. Therefore, Laue groups driven from layer groups are used to be the real-space  part of nontrivial spin group for describing 2D altermagnets. 

\begin{table}
	\renewcommand{\arraystretch}{3.8}
	\caption{\label{tab:table1}All nontrivial spin Laue groups for two dimensional altermagnets, the basic characteristic of spin-momentum locking of the materials described by these groups and their type\cite{PhysRevB.110.054406}. Here we adopt Litvin's notation of the spin groups. }
	\begin{ruledtabular}
		\begin{tabular}{ccc}
			\textrm{Type}&
			\textrm{Spin-momentum locking $(k_x,k_y)$}&
			\textrm{spin Laue group}\\ 
			\hline
			\multirow{4}*{\emph{d}-wave}&
			\multirow{2}*{\includegraphics[width=3cm]{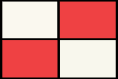}} &$^{2}2/^{2}m_x$\\ 
			& &$^{2}m^{2}m^{1}m$\\ 
			& \multirow{2}*{\includegraphics[width=2.5cm]{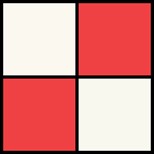}}&$^{2}4/^{1}m$\\
			& &$^{2}4/^{1}m^{2}m^{1}m$\\ 
			\multirow{2}*{\emph{g}-wave}& \multirow{2}*{\includegraphics[width=2.5cm]{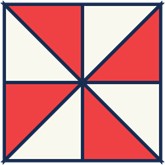}} &
			\multirow{2}*{$^{1}4/^{1}m^{2}m^{2}m$} \\
			& & \\ 
			\multirow{2}*{\emph{i}-wave}&  \multirow{2}*{\includegraphics[width=3cm]{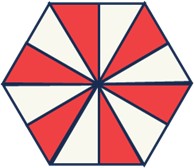}} &
			$^{1}\overline{3}$$^{2}m$ \\
			& &$^{1}6/$$^{1}m$$^{2}m$$^{2}m$ \\
		\end{tabular}
	\end{ruledtabular}
\end{table}

All nontrivial spin Laue groups describing 2D altermagnets are listed in Table \ref{tab:table1}. They are classified into three types, including \emph{d}-, \emph{g}-, \emph{i}-wave, according to the basic characteristic of spin-momentum locking. When traversing a closed loop in momentum space around the $\Gamma$-point within the \emph{xy}-plane, the spin undergoes a full 360$^\circ$ rotation through two discrete reversals and the total number of such rotations defines a characteristic even integer. When the characteristic integer is 2, 4, or 6, the corresponding types of spin-orbit coupling are respectively classified as \emph{d}-wave, \emph{g}-wave, and \emph{i}-wave\cite{PhysRevX.12.031042}, as listed in Table \ref{tab:table1}. From a symmetry perspective, the characteristic integer in 2D altermagnets is always determined by how many mirror planes perpendicular to the 2D plane or in-plane twofold rotations are accompanied by a spin rotation. For example, a \emph{d}-wave altermagnet typically hosts two such symmetries, yielding a characteristic integer of 2. The same symmetry argument applies to other types of altermagnets. Such symmetry operations correspond to spin-degenerate nodal lines passing through the $\Gamma$-point. Consequently, the type of an given altermagnet determines the appropriate $\textbf{k}$-path for both theoretical band-structure calculations and spin-resolved angle-resolved photoemission spectroscopy (SARPES) measurements, when verifying its spin splitting.
	
Different types of 2D altermagnets give rise to distinct experimentally accessible responses. The possibility of an in-plane anomalous Hall effect (AHE), where the N\'eel vector, Hall current, and electric field all lie within the 2D plane, has recently been explored in 2D altermagnets. Such an in-plane AHE is predicted to be allowed only in \emph{d}-wave altermagnets with nontrivial spin Laue group $^{2}2/^{2}m_x$ and in \emph{i}-wave altermagnets with spin Laue group $^{1}\overline{3}$$^{2}m$, while it is strictly forbidden in \emph{g}-wave altermagnets\cite{PhysRevB.111.184407}. In addition, giant magnetoresistance (GMR), derived from the anisotropic spin-momentum coupling and the anisotropic spin-dependent conductivities, is predicted to occur only in low-symmetry \emph{d}-wave altermagnets, while it is prohibited in higher-symmetry \emph{g}-wave and \emph{i}-wave altermagnets\cite{PhysRevX.12.011028}.

In 2D  altermagnets, a spin-degenerate nodal line is protected by a $[C_2\,\|\,G-H]$ symmetry when it leaves the wave vector on the line invariant. More generally, if this symmetry maps a wave vector to itself or to another vector differing by a reciprocal lattice vector, the corresponding band structure at this $\textbf{k}$-vector will exhibit spin degeneracy. In contrast,  opposite spins emerge at distinct $\textbf{k}$-vectors related by the G-H symmetry.

Symmetry plays a crucial role not only in the understanding and classification of two-dimensional altermagnets, but also in their discovery and rational design. Here, we summarize the symmetry conditions required for the emergence of two-dimensional altermagnetism. To realize two-dimensional altermagnetism, the system must host a collinear-compensated magnetic order, meaning that it contains two sublattices with antiparallel magnetic moments. The atoms belonging to these opposite sublattices must not be connected by translation symmetry, inversion symmetry, a mirror plane parallel to the xy plane, or a twofold rotation about the z axis, assuming the system lies in the \emph{xy}-plane. Additionally, there must exist at least one symmetry operation, either a proper or improper rotation, that maps one sublattice onto the other. Based on these criteria, numerous two-dimensional altermagnets have been theoretically predicted, encompassing \emph{d}-, \emph{g}-, and \emph{i}-wave types. Meanwhile, diverse approaches have emerged for realizing altermagnetism from both magnetic and nonmagnetic materials.

\section{\label{sec:level3}Survey of Two-Dimensional Altermagnets}

Theoretical efforts have identified a wide range of two-dimensional altermagnets, which span metallic, semimetallic, and semiconducting electronic properties and are predominantly characterized by \emph{d}-wave spin-momentum locking. These predicted candidates are summarized in Table \ref{tab:table2}. In terms of electronic transport properties, the majority of these materials are semiconductors. With respect to the characteristics of spin-locking, most of them exhibit \emph{d}-wave altermagnetism. Notably, only 2H-FeBr$_3$ has been identified to exhibit \emph{i}-wave altermagnetism. Under well-defined symmetry conditions for the emergence of altermagnetism, screening materials databases provides an efficient approach to discover 2D altermagnets. A systematic screening has been conducted across four databases\cite{sodequist2024two,wang2024electric,wang2025pentagonal,haddadi2025explor}: the Computational 2D Materials Database (C2DB), 2DMatPedia, Materials Cloud two-dimensional crystals database (MC2D), and Database for Pentagon-Based Sheets. 

\begin{figure*}
	
	\centering
	\includegraphics[width=\linewidth]{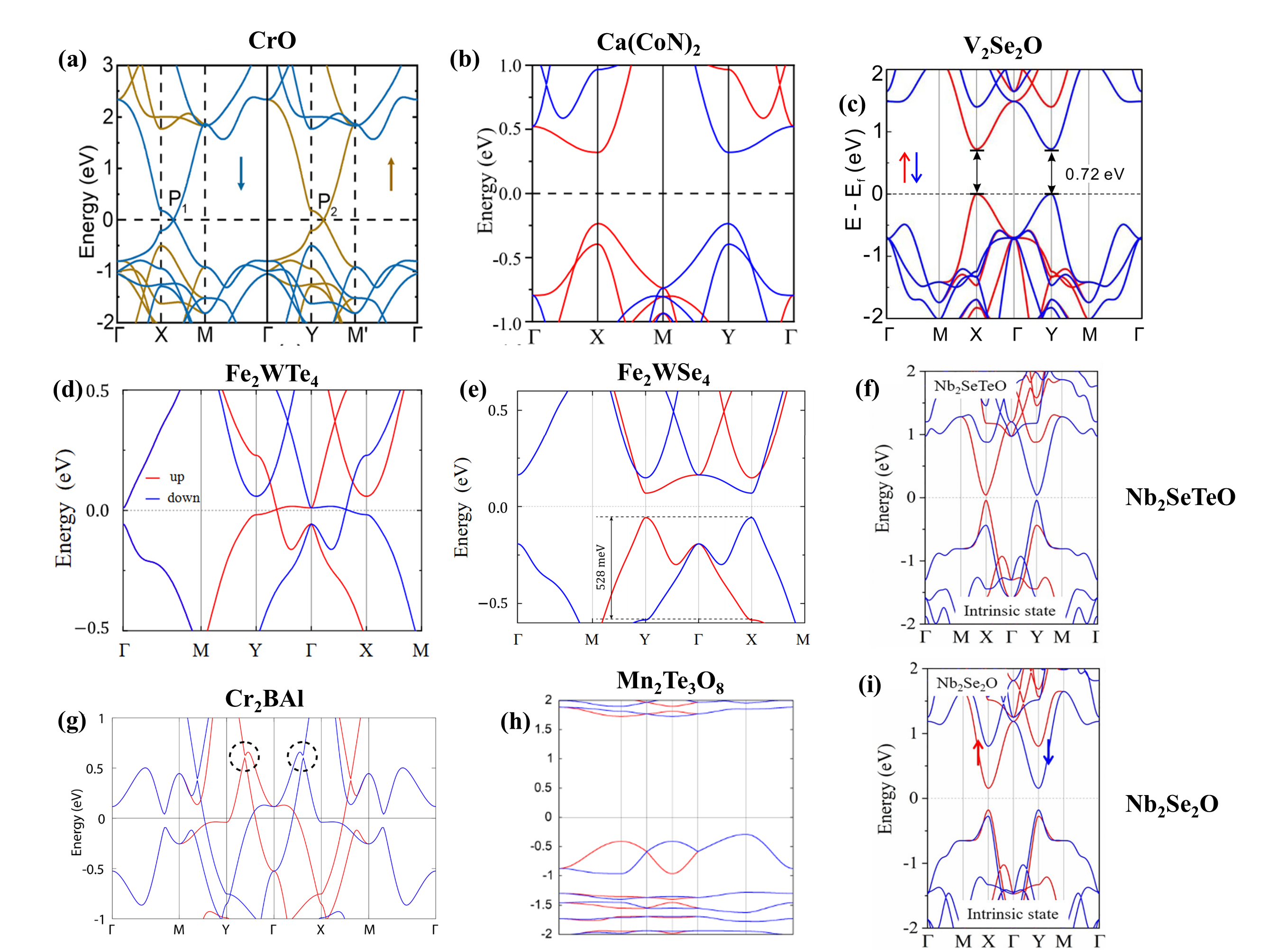}
	\caption{\label{fig:figure2}The band structure of altermagnets exhibiting large spin splitting in the absence of spin-orbit coupling: (a) CrO\cite{guo2023quantum,chen2023giant}, (b) Ca(CoN)$_2$\cite{PhysRevLett.133.056401}, (c) V$_2$Se$_2$O\cite{ma2021multifunctional}, (d) Fe$_2$WTe$_4$\cite{tan2024bipolarized}, (e) Fe$_2$WSe$_4$\cite{PhysRevB.111.094411}, (f) Nb$_2$SeTeO\cite{xie2025piezovalley}, (g) Cr$_2$BAl\cite{sattigeri2025dirac}, (h) Mn$_2$Te$_3$O$_8$\cite{wang2024electric}, (i) Nb$_2$Se$_2$O\cite{xie2025piezovalley}. All of them exhibit spin splitting greater than 500 meV.}
	
\end{figure*}

Meanwhile, several 2D functional materials exhibiting altermagnetism have also been proposed. Some materials have been identified as ferroelectric altermagnets, such as VOI$_2$\cite{zhu2025two,yang2025giant} and MgFe$_2$N$_2$\cite{guo2025altermagnetic}, where ferroelectric polarization enables the control of spin splitting in altermagnets. This mechanism has recently been proposed as a novel type of magnetoelectric coupling\cite{sun2025proposing,vsmejkal2024altermagnetic,PhysRevLett.134.106802,zhu2025two,yang2025giant}. CuMoP$_2$S(Se)$_6$ and CuWP$_2$S(Se)$_6$ have been identified as antiferroelectric altermagnets, in which a weak electric field can drive a phase transition between the antiferroelectric altermagnet and the ferroelectric antiferromagnet\cite{PhysRevLett.134.106801}. Fe$_2$WTe$_4$ and Fe$_2$MoZ$_4$ (Z = S, Te) have been identified as Weyl semimetals, which  exhibit the crystal valley Hall effect\cite{tan2024bipolarized}. V$_2$Se$_2$O is not only an altermagnet but also a novel valleytronic material, in which the valleys are connected by crystallographic symmetries rather than by time-reversal symmetry, and can therefore be modulated by external fields or geometric engineering\cite{ma2021multifunctional,PhysRevLett.124.037701,k4x8-84g8}. Building on this paradigm, a symmetry-guided chemical design framework combined with extensive high-throughput calculations has been developed, through which 670 monolayer altermagnets have been identified\cite{xuChemicalDesignMonolayer2025}. Organic materials with altermagnetism have also been investigated, such as Cr(diz)$_2$ (diz = 1,3-diazete) and \emph{t}-Cr$_2$[Pyc-O$_8$] (Pyc=pyracylene). Altermagnets with large spin splitting are more accessible to experimental observation and hold greater potential for applications. Therefore, we summarize nine altermagnets exhibiting spin splitting larger than 500 meV, and their band structures in the absence of spin-orbit coupling (SOC) are presented in Fig. \ref{fig:figure2}.

\setlength{\extrarowheight}{4pt}
\begin{longtable*}{>{\centering\arraybackslash}m{0.1\textwidth}
		>{\centering\arraybackslash}m{0.4\textwidth} 
		>{\centering\arraybackslash}m{0.15\textwidth}
		>{\centering\arraybackslash}m{0.15\textwidth}
		>{\centering\arraybackslash}m{0.12\textwidth}}
	\caption{\label{tab:table2}\parbox{1.1\linewidth}{Two-dimensional altermagnets with their types and electronic properties.} } \\
	\hline \hline
	 \multicolumn{1}{c}{\textbf{No.}} & \multicolumn{1}{c}{\textbf{Material}} & \multicolumn{1}{c}{\textbf{Type}}& \multicolumn{1}{c}{\textbf{Electronic Property}}&\multicolumn{1}{c}{\textbf{Ref.}}\\ \hline 
	\endfirsthead
	
	\multicolumn{3}{l}
	{{\bfseries \tablename\ \thetable{} -- continued from previous page}} \\
	\hline \hline
	\multicolumn{1}{c}{\textbf{No.}} & \multicolumn{1}{c}{\textbf{Material}} & \multicolumn{1}{c}{\textbf{Type}}& \multicolumn{1}{c}{\textbf{Electronic Property}}&\multicolumn{1}{c}{\textbf{Ref.}} \\ \hline 
	\endhead
	
	\hline \hline \multicolumn{3}{l}{{Continued on next page}} \\ 
	\endfoot
	
	\hline\hline
	\endlastfoot
	1&AgF$_2$&\emph{d}-wave&semiconductor&\cite{sodequist2024two}\\
	2&RuF$_4$&\emph{d}-wave&semiconductor&\cite{sodequist2024two,milivojevic2024interplay,haddadi2025explor}\\
	3&VF$_4$&\emph{d}-wave&semiconductor&\cite{sodequist2024two,haddadi2025explor}\\
	4&OsF$_4$&\emph{d}-wave&semiconductor&\cite{sodequist2024two}\\
	5&V$_2$ClBrI$_2$O$_2$&\emph{d}-wave&semiconductor&\cite{sodequist2024two}\\
	6&OsNNaSCl$_5$&\emph{d}-wave&semiconductor&\cite{sodequist2024two}\\
	7&MnTeMoO$_6$&\emph{d}-wave&semiconductor&\cite{PhysRevB.110.054406}\\
	8&Ca$_2$RuO$_4$&\emph{d}-wave&semiconductor&\cite{gonzalez2025altermagnetism}\\
	9&V$_2$Se$_2$O&\emph{d}-wave&semiconductor&\cite{ma2021multifunctional}\\
	10&CoS$_2$&\emph{d}-wave&semiconductor&\cite{wang2025pentagonal}\\
	11&CoPSe&\emph{d}-wave&semiconductor&\cite{wang2025pentagonal}\\
	12&VOX$_2$ (X=Cl,Br,I)&\emph{d}-wave&semiconductor&\cite{zhu2025two,yang2025giant}\\
	13&VX$_4$ (X=Cl,Br,I)&\emph{d}-wave&semiconductor&\cite{PhysRevB.110.155125} \\
	14&Fe$_2$WTe$_4$&\emph{d}-wave&semimetal&\cite{tan2024bipolarized}\\
	15&Fe$_2$MoZ$_4$ (Z=S,Se,Te)&\emph{d}-wave&semimetal&\cite{tan2024bipolarized}\\
	16&V$_2$Te$_2$O&\emph{d}-wave&metal&\cite{PhysRevB.111.184437,PhysRevB.108.024410}\\
	17&LiMnF$_5$&\emph{d}-wave&semiconductor&\cite{wang2024electric}\\
	18&LiVSiO$_4$&\emph{d}-wave&semiconductor&\cite{wang2024electric}\\
	19&Co$_2$Te$_3$O$_8$&\emph{d}-wave&semiconductor&\cite{wang2024electric}\\
	20&Mn$_2$Te$_3$O$_8$&\emph{d}-wave&semiconductor&\cite{wang2024electric}\\
	21&LiFeSiO$_4$&\emph{d}-wave&semiconductor&\cite{wang2024electric}\\
	22&LiMnPO$_4$&\emph{d}-wave&semiconductor&\cite{wang2024electric}\\
	23&TcF$_4$&\emph{d}-wave&semiconductor&\cite{wang2024electric}\\
	24&TcCl$_4$&\emph{d}-wave&semiconductor&\cite{wang2024electric}\\
	25&TcBr$_4$&\emph{d}-wave&semiconductor&\cite{wang2024electric}\\
	26&TcI$_4$&\emph{d}-wave&metal&\cite{wang2024electric}\\
	27&Cr$_2$SiO$_4$&\emph{d}-wave&semiconductor&\cite{wang2024electric}\\
	28&FeMgSn&\emph{d}-wave&metal&\cite{wang2024electric}\\
	29&MnMgSn&\emph{d}-wave&metal&\cite{wang2024electric}\\
	30&M(pyz)$_2$ (M = Ca and Sr, pyz = pyrazine)&\emph{d}-wave&semiconductor&\cite{che2024realizing}\\
	31&Ca(CoN)$_2$&\emph{d}-wave&semiconductor&\cite{PhysRevLett.133.056401}\\
	32&Ca(FeN)$_2$&\emph{d}-wave&semiconductor&\cite{cui2024electric}\\
	33&CrO&\emph{d}-wave&semimetal&\cite{guo2023quantum,chen2023giant}\\
	34&Cr$_2$Te$_2$O&\emph{d}-wave&semiconductor&\cite{PhysRevB.108.L180401}\\
	35&Cr$_2$Se$_2$O&\emph{d}-wave&semiconductor&\cite{PhysRevB.108.L180401}\\
	36&Fe$_2$Se$_2$O&\emph{d}-wave&semiconductor&\cite{wu2024valley}\\
	37&Cr$_2$BAl&\emph{d}-wave&metal&\cite{sattigeri2025dirac}\\
	38&Cr(diz)$_2$ (diz = 1,3-diazete)&\emph{d}-wave&semiconductor&\cite{che2025inverse}\\
	39&Cr(c-pyr)$_2$ (c-pyr = pyrrolo[3,4-c]pyrrole)&\emph{d}-wave&semiconductor&\cite{che2025inverse}\\
	40&Cr(f-pid)$_2$ (f-pid = pyrrolo[3,4-f]isoindole)&\emph{d}-wave&semiconductor&\cite{che2025inverse}\\
	41&Cr$_2$S$_2$&\emph{d}-wave&semiconductor&\cite{chen2024strain}\\
	42&CuMoP$_2$S(Se)$_6$&\emph{d}-wave&semiconductor&\cite{PhysRevLett.134.106801}\\
	43&CuWP$_2$S(Se)$_6$&\emph{d}-wave&semiconductor&\cite{PhysRevLett.134.106801}\\
	44&FeS (110)&\emph{d}-wave&semiconductor&\cite{https://doi.org/10.1002/adfm.202402080}\\
	45&FeSe (110)&\emph{d}-wave&semiconductor&\cite{https://doi.org/10.1002/adfm.202402080}\\
	46&FeSe (001)&\emph{d}-wave&metal&\cite{mazin2023induced}\\
	47&Cr$_2$SiCS$_2$&\emph{d}-wave&semiconductor&\cite{PhysRevB.111.155428}\\
	48&Cr$_2$SiCSe$_2$&\emph{d}-wave&semiconductor&\cite{PhysRevB.111.155428}\\
	49&Nb$_2$Se$_2$O&\emph{d}-wave&semiconductor&\cite{xie2025piezovalley}\\
	50&Nb$_2$SeTeO&\emph{d}-wave&semiconductor&\cite{xie2025piezovalley}\\
	51&MgFe$_2$N$_2$&\emph{d}-wave&semiconductor&\cite{guo2025altermagnetic}\\
	52&\emph{t}-Cr$_2$[Pyc-O$_8$] (Pyc=pyracylene)&\emph{d}-wave&semiconductor&\cite{v38b-5by1}\\
	53&SrRuO$_3$ (010)&\emph{d}-wave&semiconductor &\cite{samanta2020crystal} \\
	54&VSX$_2$ (X=Cl,Br,I)&\emph{d}-wave&semiconductor &\cite{zhu2025two}\\
	55&Fe$_2$WS$_4$&\emph{d}-wave&semiconductor &\cite{PhysRevB.111.094411}\\
	56&Fe$_2$WSe$_4$&\emph{d}-wave&semiconductor &\cite{PhysRevB.111.094411}\\
	57&Co$_2$MoSe$_4$&\emph{d}-wave&semiconductor&\cite{zhangMultipleStraininducedEffects2025}\\
	58&MoO&\emph{d}-wave&metal&\cite{wuQuantumAnomalousHall2023}\\
	59&MnC$_2$&\emph{g}-wave&semiconductor&\cite{wang2025pentagonal}\\
	60&MnS$_2$&\emph{g}-wave&semiconductor&\cite{wang2025pentagonal}\\
	61&Li$_2$MnP$_2$O$_7$&\emph{g}-wave&semiconductor&\cite{wang2024electric}\\
	62&VP$_2$H$_8$(NO$_4$)$_2$&\emph{g}-wave&semiconductor&\cite{PhysRevB.110.054406}\\
	63&2H-FeBr$_3$&\emph{i}-wave&semiconductor&\cite{sodequist2024two}\\
\end{longtable*}

Yet, with only a handful of two-dimensional altermagnets identified beside the 200-plus bulk crystals already catalogued \cite{adfm.202409327}, the field is clearly in its infancy. This scarcity leaves ample room---from materials discovery and property engineering to application prototyping---for the next wave of exploration. For most of the materials listed in Table \ref{tab:table2}, the magnetic ground state has been determined merely by comparing the energies of a few collinear magnetic configurations. This approach, however, is insufficient for identifying the true ground state due to  the complex magnetic energy landscape and the vast configuration space. Therefore, employing systematic and effective approaches to determine magnetic ground states, such as spin spiral calculation\cite{sodequist2024magnetic,sodequist2024two} and  Machine Learning\cite{PhysRevB.110.104427}, 
is essential for the future research on the prediction of 2D altermagnets.
\section{\label{sec:level4}Stacking Altermagnet}
Interlayer stacking---whether between identical layers or heterogeneous combinations---serves as a versatile symmetry-engineering strategy that can generate, modulate, and even switch altermagnetism in 2D systems.
Compared to three-dimensional counterparts, two-dimensional materials possess an extra 'stacking' degrees of freedom, offering a distinctive and tunable knob for symmetry engineering. Owing to the weak interlayer van der Waals (vdW) interactions, two-dimensional vdW materials can be freely stacked into either homostructures or heterostructures, thereby enabling a wide variety of physical properties\cite{wang2022magnetic}. From a symmetry perspective, different stacking configurations can result in structures with distinct crystallographic symmetries, regardless of whether they involves identical or different materials. The bilayer stacking theories provide guidance on how van der Waals monolayers with known symmetries can be stacked to yield bilayers with desired crystallographic symmetries\cite{PhysRevLett.130.146801}. This establishes the theoretical basis for rationally designing 2D systems with targeted symmetries, while ongoing advances in precise stacking techniques now make their laboratory assembly an experimental reality\cite{frisenda2018recent,son2020strongly,liu2016van,kinoshita2019dry}. Because altermagnetism is symmetry-governed\cite{PhysRevB.110.054406,PhysRevX.12.031042}, layer stacking becomes a readily accessible experimental lever for creating it in two dimensions. Here we survey practical stacking protocols---using identical or distinct 2D magnets---and highlight the device opportunities they unlock.

Although experimental demonstrations of altermagnetism realized through stacking in 2D materials are still absent, existing fabrication and assembly techniques\cite{frisendaRecentProgressAssembly2018,qiProductionMethodsVan2018,liu2016van,songAdvancedSynthesisInfluencing2025} clearly show that this approach is experimentally feasible. Traditional approaches to constructing stacked 2D bilayer or multilayer can be grouped broadly into two categories: 'top-down' transfer and assembly, and 'bottom-up' direct growth. 'Top-down' methods first produce monolayers by mechanical or chemical exfoliation and then stack them using an appropriate transfer technique\cite{yangSituManipulationVan2020,ceballosUltrafastChargeSeparation2014}. Common transfer strategies include dry-transfer\cite{kinoshita2019dry, TransferTwodimensional2014} and wet-transfer\cite{schneiderWedgingTransferNanostructures2010}. This method allow flexible and precise stacking of dissimilar layers, enabling controlled layer sequence, relative twist angle, and atomically clean heterointerfaces. On the other hand, 'bottom-up' direct growth techniques, such as chemical vapor deposition (CVD) and molecular beam epitaxy (MBE), avoids contamination or damage associated with transfer steps and is better suited to large-area, scalable fabrication\cite{zhouStackGrowthWaferscale2023,liuDesignedGrowthLarge2022}. Notably, wafer-scale, uniform bilayer and multilayer 2D materials (e.g., graphene and various transition-metal dichalcogenides) have already been successfully synthesized\cite{liuDesignedGrowthLarge2022,kangLayerbylayerAssemblyTwodimensional2017,zhouStackGrowthWaferscale2023}. Taken together, although stacked 2D altermagnets have not yet been realized experimentally, the existing stacking technologies for 2D materials, ranging from finely controlled 'top-down' assembly to scalable 'bottom-up' growth, are already well established, providing a solid technological foundation for their future experimental implementation.

\subsection{\label{Bilayer}Homostructure Stacking}
\begin{figure*}
	
	\centering
	\includegraphics[width=\linewidth]{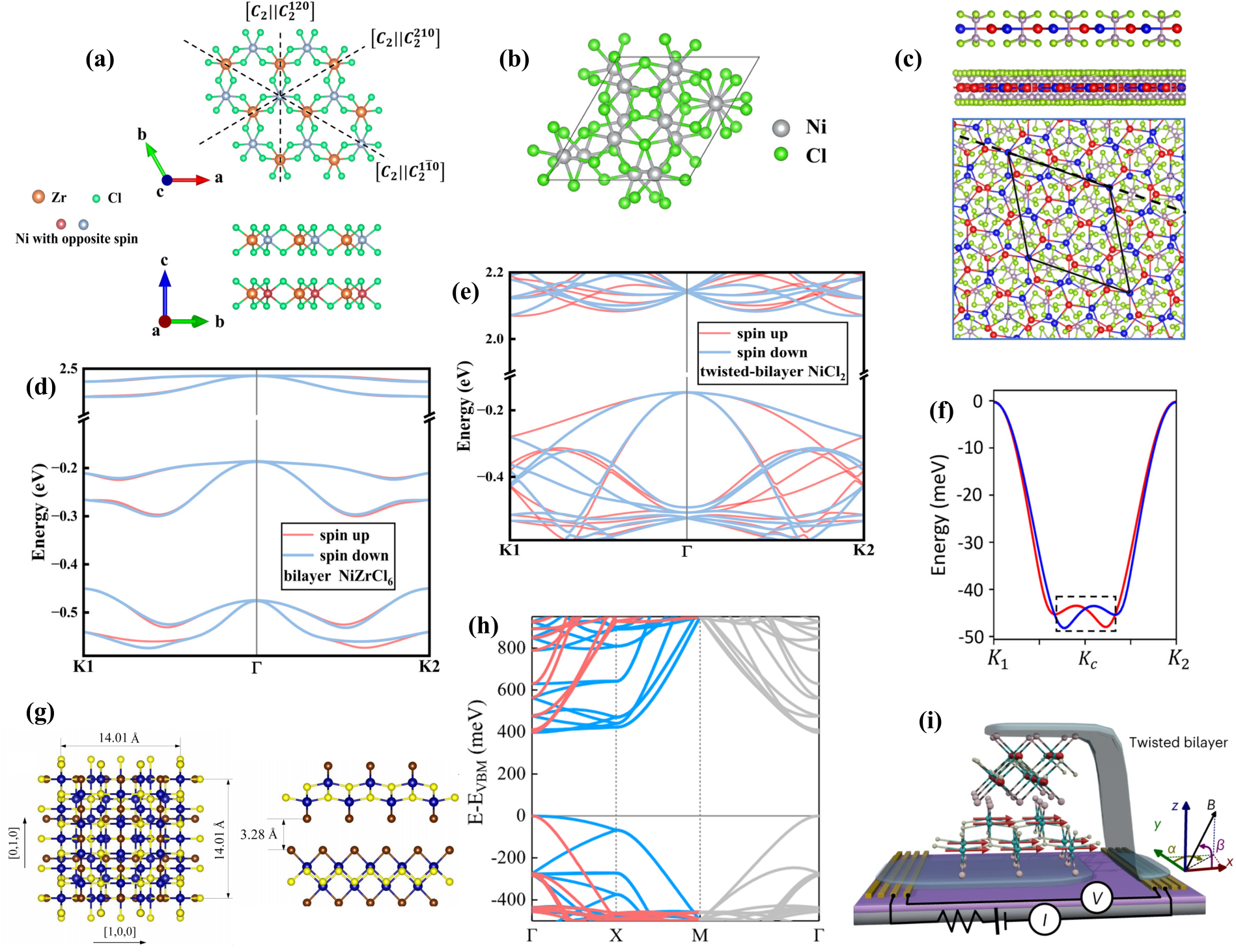}
	\caption{\label{fig:figure3.1}The crystal structure and band structure without spin-orbit coupling (SOC) of bilayer altermagnets. (a)-(c) and (g) The crystal structure of bilayer NiZrCl$_6$\cite{PhysRevB.110.174410}, twisted NiCl$_2$ (with a twisting angle of $21.79^\circ$)\cite{PhysRevB.110.174410}, twisted MnPSe$_3$ (with a twisting angle of $21.79^\circ$)\cite{PhysRevMaterials.8.L051401} and twisted CrSBr (with a twisting angle of $90^\circ$)\cite{PhysRevLett.130.046401}, respectively. (d)-(f) and (h) The band structure without SOC of bilayer NiZrCl$_6$, twisted NiCl$_2$ (with a twisting angle of $21.79^\circ$), twisted MnPSe$_3$ (with a twisting angle of $21.79^\circ$) and twisted CrSBr (with a twisting angle of $90^\circ$) respectively. (i) The orientations of magnetic moments in the orthogonal bilayer  CrSBr\cite{boix2024multistep}. Red arrows denote the spin orientations along the easy magnetic axis, assuming that interlayer magnetic interactions are negligible.}
	
\end{figure*} 
The realization of a bilayer altermagnet in a homostructure can be achieved either by stacking two antiferromagnetic layers or by stacking two ferromagnetic layers with interlayer antiferromagnetic coupling, thereby ensuring a zero net magnetic moment. In previous studies, twisted bilayer altermagnets have been proposed\cite{PhysRevLett.130.046401,PhysRevMaterials.8.L051401}, which can be regarded as stacking altermagnets, since twisting constitutes a particular stacking operation. In addition, several special stacking configurations have been proposed, including the flipped\cite{PhysRevLett.133.206702} and reversed\cite{PhysRevB.110.224418} types. Moreover, an approach was proposed to systematically derive the stacking configurations of bilayer A-type altermagnets\cite{PhysRevB.110.174410,PhysRevLett.133.166701}. Recent studies in multiferroics have proposed novel magnetoelectric coupling mechanisms, where sliding ferroelectricity is employed to modulate spin splitting\cite{c4dm-12ng,sun2025proposing,zhu2025sliding}. In this subsection, we outline the derivation of bilayer A-type altermagnets, introduce representative examples of 2D altermagnets formed by homostructure stacking, and discuss their potential applications. 

We first introduce the bilayer A-type altermagnet, characterized by ferromagnetic coupling within each layer and antiferromagnetic coupling between layers, i.e., an altermagnet realized through stacking ferromagnetic monolayers\cite{PhysRevB.110.174410,PhysRevLett.133.166701}. In A-type altermagnets, sublattices with opposite spins cannot be connected by translational symmetry, a consequence of the intrinsic absence of out-of-plane translational symmetry in 2D materials. Moreover, since the intralayer magnetic moments are aligned in parallel, they also cannot be connected by a twofold rotational symmetry around the \emph{z}-axis. According to Section \ref{sec:level2}, the spin-degenerate band structure is protected by the symmetry operations $[C_2\,\|\,A]$, where $A \in \{\mathcal{P}, \tau, m_z, C_{2z}\}$. Therefore, the emergence of A-type altermagnetism merely requires the breaking of $[C_2\,\|\,\mathcal{P}]$ and $[C_2\,\|\,m_z]$. Therefore, A-type altermagnets are more likely to emerge.

Owing to the nonrelativistic nature of altermagnetism, real space and spin space are decoupled. Since the magnetic order in A-type altermagnets is predetermined, the symmetry analysis is confined to real-space operations alone\cite{PhysRevB.110.174410}. Therefore, the stacking model, originally proposed in the field of ferroelectrics\cite{PhysRevLett.130.146801}, can also be applied in the present context. The stacking model can be formulated as follows:
\begin{equation}
	\left\{\label{eq:5}
	\begin{aligned}
		S &= \hat{R}_sS, \ \ \forall \hat{R}_s \in G_s \\
		B &= \hat{R}_BB, \ \ \forall \hat{R}_B \in G_B\\
		B &= S+S' \\
		S'&= \hat{\tau}_z\hat{O} S
	\end{aligned}
	\right.
\end{equation}
Here, $S$ denotes a single ferromagnetic layer with layer group $G_s=\{\hat{R}_s\}$, where $\hat{R}_s$ represents the space-group symmetry operation. The second ferromagnetic layer $S'$ is generated by applying the stacking operation $\hat{\tau}_z\hat{O}$ on $S$, where $\hat{\tau}_z$ is an arbitrary trivial translation operation along \emph{z} direction and cannot affect the symmetry of the bilayer $B$ whose layer group is $G_B=\{\hat{R}_B\}$. $\hat{O}$ is the non-trivial part of stacking operation. The bilayer $B$ is composed of two ferromagnetic layers, $S$ and $S'$.

The emergence of altermagnetism is primarily determined by the symmetry operations between sublattices, as discussed in Section \ref{sec:level2}, specifically the interlayer symmetry operations in A-type stacking. Therefore, it is sufficient to consider only the symmetries of the stacked bilayer that interchange the two layers. The complete set of 2D point-group symmetry operations capable of reversing the \emph{z}-coordinate can be expressed as $\{\overline{E},m_z,C_{2\alpha},S_{nz}\}$. For a ferromagnetic layer with a known layer group, the stacking model can be used to determine the stacking operations required to generate or preserve each of the symmetry operations $\{\overline{E},m_z,C_{2\alpha},S_{nz}\}$ in the stacked bilayer. By further excluding the stacking operations that yield $\{\overline{E},m_z\}$ from those giving rise to $\{C_{2\alpha},S_{nz}\}$, one ensures that the sublattices in the stacked bilayer are not connected by $\overline{E}$ or $m_z$, while still being related by at least one symmetry operation. Through the above procedure, all stacking operations capable of generating 2D altermagnetism can be identified for a given monolayer. A detailed derivation is available in Ref.\cite{PhysRevB.110.174410}.
 
In addition, an alternative approach has also been proposed\cite{PhysRevLett.133.166701}. For a bilayer composed of monolayers $S$ and $S'$, the symmetry operations can be divided into two sets. The intersection set
\begin{align}
	\label{eq:6}
	Z_i = G_s \cap G_{s'}
\end{align}
contains the symmetry elements common to both monolayers, where $G_s$ denotes the layer group of monolayer $S$, and $G_{s'}$ is obtained by applying the stacking operator $\hat{O}$ to $G_s$, i.e., $G_{s'} = \hat{O} G_s \hat{O}^{-1}$. The exchange set 
\begin{align}
\label{eq:7}
Z_e = \hat{O} G_s \cap G_s \hat{O}^{-1}
\end{align}
consists of operations that interchange $S$ and $S'$. The symmetry group of the bilayer is then given by 
\begin{align}
\label{eq:8}
G_B = Z_i \cup Z_e,
\end{align} 
which can be determined directly from the monolayer layer group $G_s$ and the stacking operator $\hat{O}$. By considering all symmetry operations of the monolayer and the stacking operator $\hat{O}$, the symmetry of the bilayer under different stacking configurations can be obtained, and subsequently examined to determine whether the conditions for 2D altermagnetism are satisfied. A more detailed discussion can be found  Ref.\cite{PhysRevLett.133.166701}.

Based on the above theories, the main conclusions regarding A-type altermagnets are drawn and summarized as follows. Intrinsic A-type altermagnetism, which arises from the direct stacking of two identical monolayers, can only be realized in 17 specific layer groups (8-10, 19, 22, 50, 53-54, 57-60, 67-68, and 76). Accordingly, A-type antiferromagnetic bilayers belonging to these groups inherently exhibit altermagnetism. The sublattices of A-type altermagnets can be connected only through $S_{4z}$ or $C_{2\alpha}$ operations. Moreover, a $C_{2\alpha}$ rotation serves as a general stacking operation capable of generating an A-type altermagnet from an arbitrary monolayer. In addition, if the initial monolayer possesses an in-plane $C_2$ rotation symmetry, then its twisted bilayer with a twist angle $\theta$ must possess an in-plane $C_2$ axis. In this case, the angle between the $C_2$ axes of the monolayer and the twisted bilayer is $\theta/2$, which generally preserves altermagnetism in the bilayer.

Several bilayer altermagnets are illustrated in Fig. \ref{fig:figure3.1}. Panels (a) and (b) show the crystal structures of bilayer NiZrCl$_6$ and NiCl$_2$, respectively, the latter possessing a twisting angle of $21.79^\circ$. Both systems exhibit A-type altermagnetism\cite{PhysRevB.110.174410}. In these systems, the spin sublattices are connected by an in-plane $C_2$ symmetry. The corresponding nonrelativistic band structures of bilayer NiZrCl$_6$ and NiCl$_2$ are presented in Figs. \ref{fig:figure3.1}(d) and (e), indicating that bilayer NiCl$_2$ exhibits a more pronounced spin splitting compared with bilayer NiZrCl$_6$. We next provide some qualitative insights to understand this difference. Owing to the weak vdW interaction between layers, the band structure of a bilayer can be approximated as simply composed of the electronic structures of its constituent monolayers. For each layer of an A-type altermagnet, when considered individually, the spin-up (spin-down) band structure of the top layer is identical to the spin-down (spin-up) band structure of the bottom layer. Because bilayer NiZrCl$_6$ is formed by direct stacking, it exhibits only a very small spin splitting. This splitting originates from perturbations induced by interlayer vdW interactions, whose influence on the band structure is not significant. This point is further supported by the fact that the spin splitting in bilayer NiZrI$_6$ increases with the strength of the van der Waals interactions. In contrast, bilayer NiCl$_2$ is obtained by rotating the top layer by $21.79^\circ$ before stacking, which leads to a misalignment between the band structures of the two layers and consequently results in a large spin splitting. Furthermore, if the bandwidth of the monolayer is narrow, the spin splitting in the bilayer will be small regardless of the stacking configuration. Therefore, for A-type altermagnets, a larger bandwidth of band structure  in the monolayer generally favors a more pronounced spin splitting in the bilayer. Fig. \ref{fig:figure3.1}(h) reveals that $90^\circ$-twisted bilayer CrSBr---whose crystal structure is depicted in Fig. \ref{fig:figure3.1}(g)---displays a pronounced spin splitting\cite{PhysRevLett.130.046401}. This arises not only from the large bandwidth but also from the strong anisotropy of the band structure of the monolayer CrSBr. Therefore, the magnitude of spin splitting in bilayer A-type altermagnets is governed by two key factors: the bandwidth and the anisotropy of the band structure in the monolayers. 

Although altermagnetism is a nonrelativistic effect and studies on altermagnets usually neglect SOC, this approximation appears to be problematic for bilayer CrSBr. As is well known, CrSBr exhibits strong in-plane magnetocrystalline anisotropy originating from the SOC effect\cite{lee2021magnetic}. When constructing a twisted bilayer of CrSBr with a $90^\circ$ rotation, the \emph{a} axis of one layer is aligned parallel to the \emph{b} axis of the other, rendering the easy magnetic axes of the two layers mutually perpendicular and resulting in a coplanar magnetic configuration, as illustrated in Fig. \ref{fig:figure3.1}(i)\cite{boix2024multistep}. Such an insight appears to be in conflict with the previous report claiming that bilayer CrSBr is an altermagnet\cite{PhysRevLett.130.046401}. Therefore, the impact of SOC on this system should be further investigated, and whether it qualifies as an altermagnet remains an open question. Furthermore, this issue generally arises when stacking 2D materials with an in-plane spin orientation.

In addition to A-type altermagnets, we present a bilayer of the $\mathcal{PT}$ antiferromagnet MnPSe$_3$, stacked with a twisting angle of $21.79^\circ$, which exhibits 2D altermagnetism, as shown in Fig. \ref{fig:figure3.1}(c)\cite{PhysRevMaterials.8.L051401}. Because each layer in this bilayer is a spin-degenerate antiferromagnet, the spin splitting originates mainly from the perturbations induced by interlayer vdW interactions and is therefore very small, as illustrated in Fig. \ref{fig:figure3.1}(f). From a qualitative perspective, 2D altermagnets constructed from spin-degenerate antiferromagnets are not expected to exhibit pronounced spin splitting.

\begin{figure*}
	
	\centering
	\includegraphics[width=\linewidth]{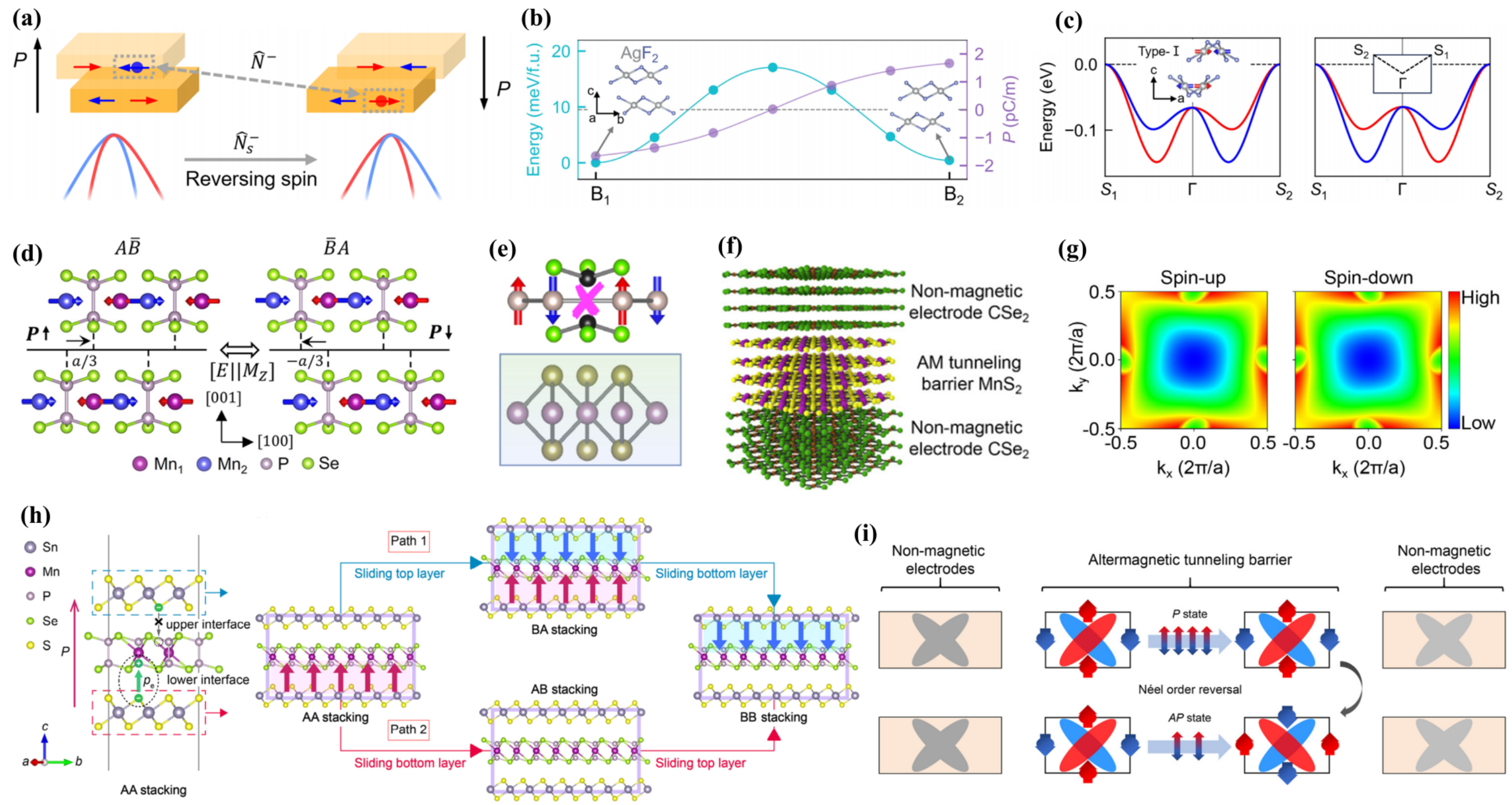}
	\caption{\label{fig:figure3.2}(a) Schematic diagram of sliding-ferroelectricity-controlled spin-splitting switching, including configurations (upper) and bands (lower)\cite{zhu2025sliding}. (b) Energy barrier and evolution of out-of-plane polarization along the sliding path between AgF$_2$ bilayers B1 and B2, with the inset showing the corresponding B1 and B2 configurations. B1 and B2 are related by the $[C_2\,\|\,M_z]$ symmetry\cite{zhu2025sliding}. (c) The band structures of B1 (left) and B2 (right) without spin-orbit coupling\cite{zhu2025sliding}. (d) Side view of the MnPSe$_3$ bilayer in the ground states $A\overline{B}$ (left) and $\overline{B}A$ (right), which are related by the $[E\,\|\,M_z]$ symmetry\cite{c4dm-12ng}. (e) The crystal structure of heterostructure MnPSe$_3$/MoTe$_2$\cite{mazin2023induced}. (f) Device configuration of the altermagnetic tunneling junction (AMTJ) CSe$_2$/MnS$_2$/CSe$_2$\cite{wang2025pentagonal}. (g) Lowest decay rates of MnS$_2$ evanescent states along the [001] direction, resolved with respect to spin and in-plane momentum ($k_{\parallel}$)\cite{wang2025pentagonal}. (h) Stacking configuration of the trilayer heterostructure SnS$_2$/MnPSe$_3$/SnS$_2$ and two ferroelectric transition paths in the heterostructure\cite{sun2024altermagnetism}. (i) Mechanism and working states of the two-dimensional AMTJ, with red and blue representing opposite spins\cite{wang2025pentagonal}.}

\end{figure*} 

In the field of ferroelectrics, an interesting phenomenon has attracted considerable attention: when two nonpolar vdW layers are stacked in a noncentrosymmetric configuration, an out-of-plane electric polarization emerges, which can be reversed via interlayer sliding\cite{doi:10.1126/science.abi7296,doi:10.1073/pnas.2115703118}. This phenomenon is referred to as sliding ferroelectricity. Compared with conventional ferroelectrics, sliding ferroelectrics offer notable advantages, such as enhanced fatigue resistance and low energy consumption, rendering them highly promising for next-generation spintronic applications\cite{zhang2025emerging}. Stacking altermagnets provide a promising platform for exploring the interplay between sliding ferroelectricity and altermagnetism, paving the way toward the realization of unconventional magnetoelectric couplings. Recent studies have primarily focused on the coupling between the nonrelativistic spin splitting and sliding ferroelectricity in stacking altermagnets, as shown in Fig. \ref{fig:figure3.2}(a)\cite{zhu2025sliding,c4dm-12ng,sun2024altermagnetism}. 

To achieve the coexistence of altermagnetism and ferroelectricity, the stacked configuration must fulfill the symmetry criteria of both. Specifically, the real-space part of the spin group of a stacking altermagnet should correspond to a polar group. In such systems, two distinct cases arise: (i) the nonrelativistic spin splitting remains unchanged upon ferroelectric polarization reversal, and (ii) the spin splitting reverses concomitantly with the ferroelectric polarization. From a symmetry perspective, the two opposite ferroelectric states connected by sliding can be regarded as being related through a specific spin-group operation. The influence of this operation on the spin texture determines which case the stacking altermagnet belongs to. For example, stacking configurations $\{m_{001}|(0.5,-0.16)\}$ and $\{m_{001}|(0.5,0.16)\}$ correspond to two opposite ferroelectric states of bilayer AgF$_2$, which are related by the spin-group operation $[C_2||m_z]$, as shown in Fig. \ref{fig:figure3.2}(b)\cite{zhu2025sliding}. This operation reverses the spin splitting, i.e., $[C_2||m_z]\varepsilon_1(s,\textbf{k})=\varepsilon_2(-s,\textbf{k})$, where $\varepsilon_1$ and $\varepsilon_2$ denote the energies of the two distinct stacking configurations. Consequently, the nonrelativistic spin splitting reverses in tandem with the reversal of ferroelectric polarization, as illustrated in Fig. \ref{fig:figure3.2}(c). For bilayer MnPSe$_3$, the $A\overline{B}$ ($\{m_{001}|(\frac{1}{3},0)\}$) and $\overline{B}A$ ($\{m_{001}|(-\frac{1}{3},0)\}$) stacking configurations correspond to two opposite ferroelectric states, which are related by the $[E||m_z]$ symmetry operation, as shown in Fig. \ref{fig:figure3.2}(d)\cite{c4dm-12ng}. This operation unchange the spin splitting, i.e., $[E||m_z]\varepsilon_{A\overline{B}}(s,\textbf{k})=\varepsilon_{\overline{B}A}(s,\textbf{k})$. Therefore, sliding between the two ferroelectric states leads to a reversal of polarization, whereas the spin splitting remains invariant.

Although significant progress has been made in exploring bilayer altermagnets, existing studies have largely focused on how stacking, twisting, or sliding can generate altermagnetism, while the underlying interlayer exchange-coupling mechanisms remain poorly understood. Drawing insights from previous work on conventional magnetic bilayers, such as the Bethe-Slater-curve-like behavior reported in transition-metal dichalcogenide bilayers\cite{wangBetheSlatercurvelikeBehaviorInterlayer2020}, it is reasonable to expect that similar interlayer-coupling phenomena may also emerge in 2D altermagnets. In those systems, the competition between Pauli (Coulomb) repulsion and kinetic-energy gain across the van der Waals gap determines whether the interlayer interaction favors ferromagnetic or antiferromagnetic alignment. Such mechanisms, if present in altermagnets, would enable fine control of the overall magnetic state through vertical pressure or stacking engineering, offering an additional degree of freedom for designing multifunctional spintronic devices. Looking ahead, elucidating the interlayer exchange pathways, the dominant interaction mechanisms, and their dependence on structural degrees of freedom in bilayer altermagnets will open a new avenue for achieving highly tunable altermagnetic states.
\subsection{\label{Heterostructure}Heterostructure Stacking} 
In addition to homostructure stacking, heterostructure stacking can also break the symmetry, thereby transforming an antiferromagnet into an altermagnet. Evidently, bilayer heterostructure stacking breaks both spatial inversion and mirror symmetry with respect to the $xy$ plane, thereby lifting the spin degeneracy. For example, MnPSe$_3$, a $\mathcal{PT}$ antiferromagnet, can acquire altermagnetism when stacked with MoTe$_2$ to form a heterostructure that breaks $\mathcal{PT}$ symmetry, as shown in Fig. \ref{fig:figure3.2}(e)\cite{mazin2023induced}. In contrast, trilayer stacking is more intricate, as it does not inherently break any specific symmetry. For instance, the trilayer heterostructure SnS$_2$/MnPSe$_3$/SnS$_2$ retains $\mathcal{PT}$ symmetry under both AB and BA stacking\cite{sun2024altermagnetism}. In the AA and BB stacking configurations, the breaking of $\mathcal{PT}$ symmetry induces both ferroelectric polarization and altermagnetism in the heterostructure, with the polarization primarily arising from interlayer charge transfer between S and P atom pairs, as shown in Fig. \ref{fig:figure3.2}(h). These two stacking configurations possess opposite ferroelectric polarizations. They are related by the $\mathcal{PT}$ operation, which in turn results in the two ferroelectric states exhibiting opposite spin splittings. Therefore, when sliding from the AA to the BB stacking configuration, both the ferroelectric polarization and the nonrelativistic spin splitting are simultaneously reversed. Because both the upper and lower SnS$_2$ monolayers can slide relative to the MnPSe$_3$ layer, the heterostructure exhibits two transition pathways for switching the polarization from AA to BB stacking, as shown in Fig. \ref{fig:figure3.2}(h). From the perspective of the energy barrier, Path 1 is more favorable for polarization switching.

From an application perspective, heterostructures can also be employed to realize giant magnetoresistance or tunneling magnetoresistance\cite{wang2025pentagonal}. An altermagnetic tunneling junction (AMTJ) is illustrated in Fig. \ref{fig:figure3.2}(f), where the altermagnet MnS$_2$ serves as the insulating layer and the nonmagnetic metal CSe$_2$ forms the electrodes on both sides. In MnS$_2$, $k_{\parallel}$-dependent evanescent states have spin-dependent decay rates, as illustrated in Fig. \ref{fig:figure3.2}(g). Therefore, in the parallel (P) state, where all layers share the same N\'eel orientation, spin-up and spin-down carriers tunnel through relatively low barriers, yielding a large current, while reversing the N\'eel order in half of the altermagnetic insulating layers (AP state) drastically suppresses the current. A schematic illustration of the AMTJ mechanism is shown in Fig. \ref{fig:figure3.2}(i). A giant TMR of $1.5 \times 10^{5}\%$ at the Fermi level is predicted in the AMTJ CSe$_2$/MnS$_2$/CSe$_2$. In addition, unconventional proximity effects have been explored in heterostructures composed of altermagnets and nonmagnetic materials, where the characteristic momentum-alternating spin splitting of the altermagnet is directly transferred to the adjacent nonmagnetic layers\cite{zhu2025altermagneticproximityeffect}.

\section{\label{sec:level5}Multicomponent altermagnet}
\begin{figure*}
	
	\centering
	\includegraphics[width=\linewidth]{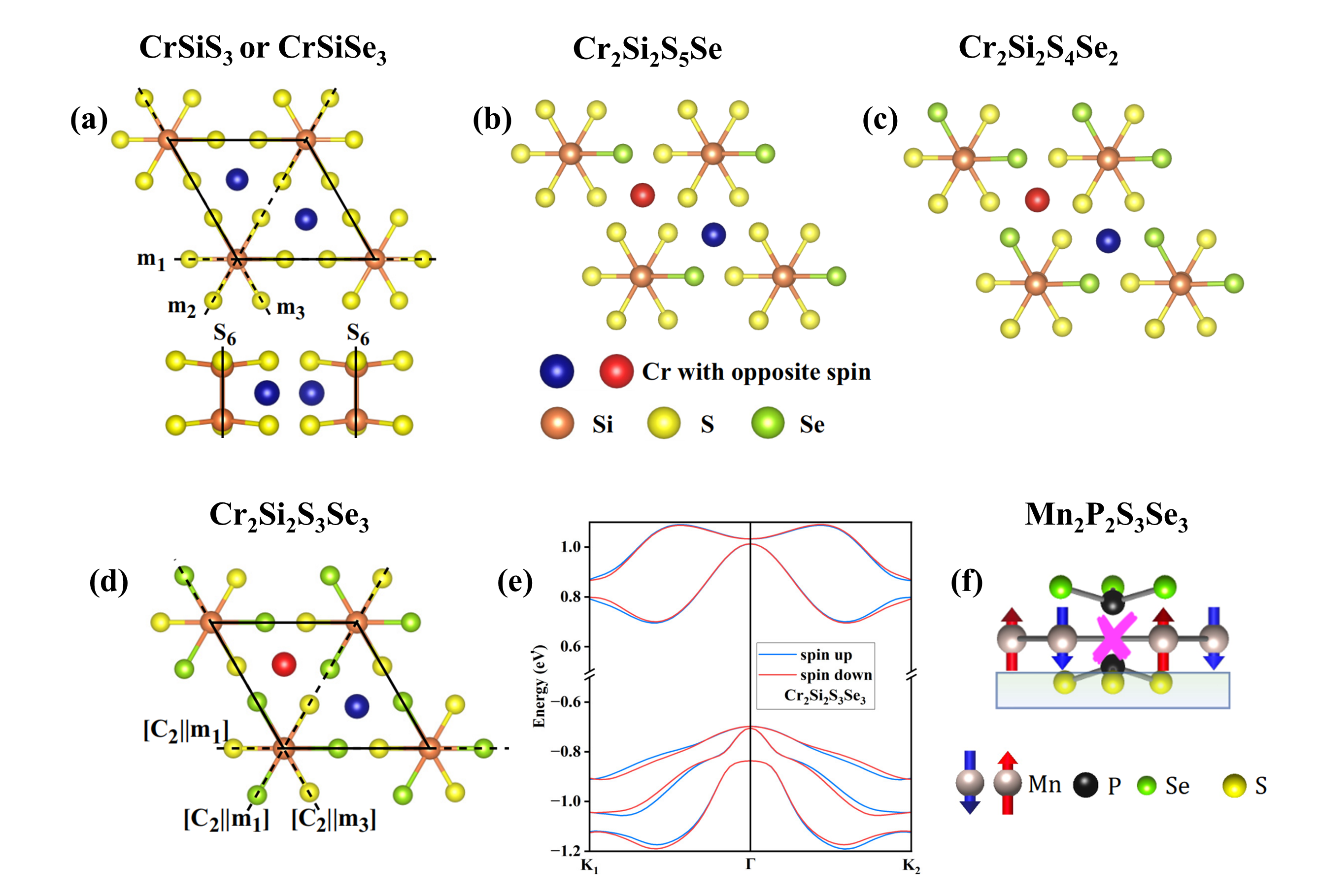}
	\caption{\label{fig:figure4}(a) The crystal structure of CrSiS$_3$ or CrSiSe$_3$\cite{PhysRevB.111.195123}. (b) The crystal structure of Cr$_2$Si$_2$S$_5$Se, which is a multicomponent derivative formed from CrSiS$_3$ and CrSiSe$_3$ with a sulfur-to-selenium ratio of 5:1\cite{PhysRevB.111.195123}. (c) and (d) The crystal structure of Cr$_2$Si$_2$S$_4$Se$_2$ and Cr$_2$Si$_2$S$_3$Se$_3$\cite{PhysRevB.111.195123}. (e) The band structure of Cr$_2$Si$_2$S$_3$Se$_3$ without spin-orbit coupling\cite{PhysRevB.111.195123}. (f) The crystal structure of Mn$_2$P$_2$S$_3$Se$_3$\cite{mazin2023induced}.}
	
\end{figure*}
Chemical alloying and compositional asymmetry emerge as a potent strategy for unlocking altermagnetism in two-dimensional multicomponent systems, where they enable precise control over symmetry breaking. Multicomponent altermagnets are a type of altermagnet constructed by combining two materials with the same symmetry and similar chemical composition, following the alloying concept\cite{PhysRevB.111.195123}. As discussed in Section \ref{sec:level2}, the spin-degenerate band structure in spin-degenerate antiferromagnets is protected by the four symmetry operations $[C_2\,\|\,A]$, where $A \in \{\mathcal{P}, \tau, m_z, C_{2z}\}$.By constructing multicomponent structures from spin-degenerate antiferromagnets and excluding these symmetry operations, one can generate candidate structures from which a wide range of altermagnets can be identified. 

For materials with similar element and crystal structure, a series of rich multicomponent structures can be obtained by mixing two or more components in different concentrations. For example, CrSiS and CrSiSe share the same crystal structure and have similar chemical compositions, as illustrated in Fig. \ref{fig:figure4}(a). Based on this crystal structure, multicomponent compounds can be constructed by varying the concentrations of S and Se, as shown in Figs. \ref{fig:figure4}(b)-(d). Specifically, Cr$_2$Si$_2$S$_5$Se corresponds to an S:Se ratio of 5:1, Cr$_2$Si$_2$S$_4$Se$_2$ corresponds to 2:1, and Cr$_2$Si$_2$S$_3$Se$_3$ corresponds to 1:1. These multicomponent compounds may have different symmetries. Through symmetry analysis, candidate structures that satisfy the symmetry requirement of altermagnetism can be screened out. Furthermore, by calculating the band structure to verify their spin splitting, a series of multicomponent altermagnets can be obtained. The band structure of Cr$_2$Si$_2$S$_3$Se$_3$, calculated without SOC, is presented in Fig. \ref{fig:figure4}(e).

Janus engineering has emerged as a popular strategy for obtaining altermagnets\cite{mazin2023induced,liu2025realizing,ma2025multifunctional,singh2025v2se2o,guo2023piezoelectric,shvd-vmvs}. In fact, the altermagnets realized through this approach constitute a subset of multicomponent altermagnets. In Janus structures, both spatial inversion and mirror symmetry in the xy-plane are broken. As a result, the spin degeneracy protected by the combination of these operations with spin reversal is lifted, thereby offering a route to realize altermagnetism through Janus engineering. For example, MnPSe$_3$ is a two-dimensional antiferromagnetic semiconductor whose spin degeneracy is protected by $\mathcal{PT}$ symmetry\cite{PhysRevB.94.184428}. Replacing one Se layer with S yields the Janus structure Mn$_2$P$_2$S$_3$Se$_3$, as shown in Fig. \ref{fig:figure4}(f), in which the breaking of $\mathcal{PT}$ symmetry results in a nonrelativistic spin-split band structure\cite{mazin2023induced}. It can also be regarded as a multicomponent altermagnet formed by combining MnPS$_3$ and MnPSe$_3$ with an S:Se ratio of 1:1. Janus engineering can also be applied to extend 2D altermagnets based on the reported altermagnets. For example, V$_2$SeTeO, a Janus derivative of V$_2$Se$_2$O and V$_2$Te$_2$O, exhibits 2D altermagnetism\cite{singh2025v2se2o}.

To enable experimental synthesis, it is crucial to evaluate the stability of the proposed crystals. A portion of the multicomponent structures generated by the aforementioned methods may lack stability and, as a result, may be challenging to realize experimentally. To assess dynamic stability, it is necessary to analyze the phonon spectra of multicomponent altermagnets. A structure is deemed dynamically stable if its phonon spectrum exhibits no imaginary frequencies. For thermodynamic stability, the formation energy should be calculated. Assuming that in the multicomponent material, A is a fixed component, B and C are adjustable components, and the multi-component material is AB$_x$C$_{(1-x)}$, the formation enthalpy is calculated by
\begin{equation*}
	\label{eq:9}
	\Delta H_F = E(AB_x C_{(1-x)}) - xE(AB) - (1-x)E(AC)
\end{equation*}
in which $E(AB_x C_{(1-x)})$ denotes the total energy of AB$_x$C$_{(1-x)}$ obtained from density functional theory (DFT) calculations and $E(AB)$ and $E(AC)$ the corresponding total energies of the binary compounds. If the formation enthalpy of a structure is less than 0 and, at the same time, is the smallest among all multicomponent structures at the same concentration, then the multicomponent structure AB$_x$C$_{(1-x)}$ is thermodynamically stable, and the phase separation fails to appear. Previous studies have predominantly focused on dynamic stability, whereas the evaluation of thermodynamic stability is equally essential.

\section{\label{sec:level6}Achieving Altermagnetism via Adsorption}
\begin{figure*}
	
	\centering
	\includegraphics[width=\linewidth]{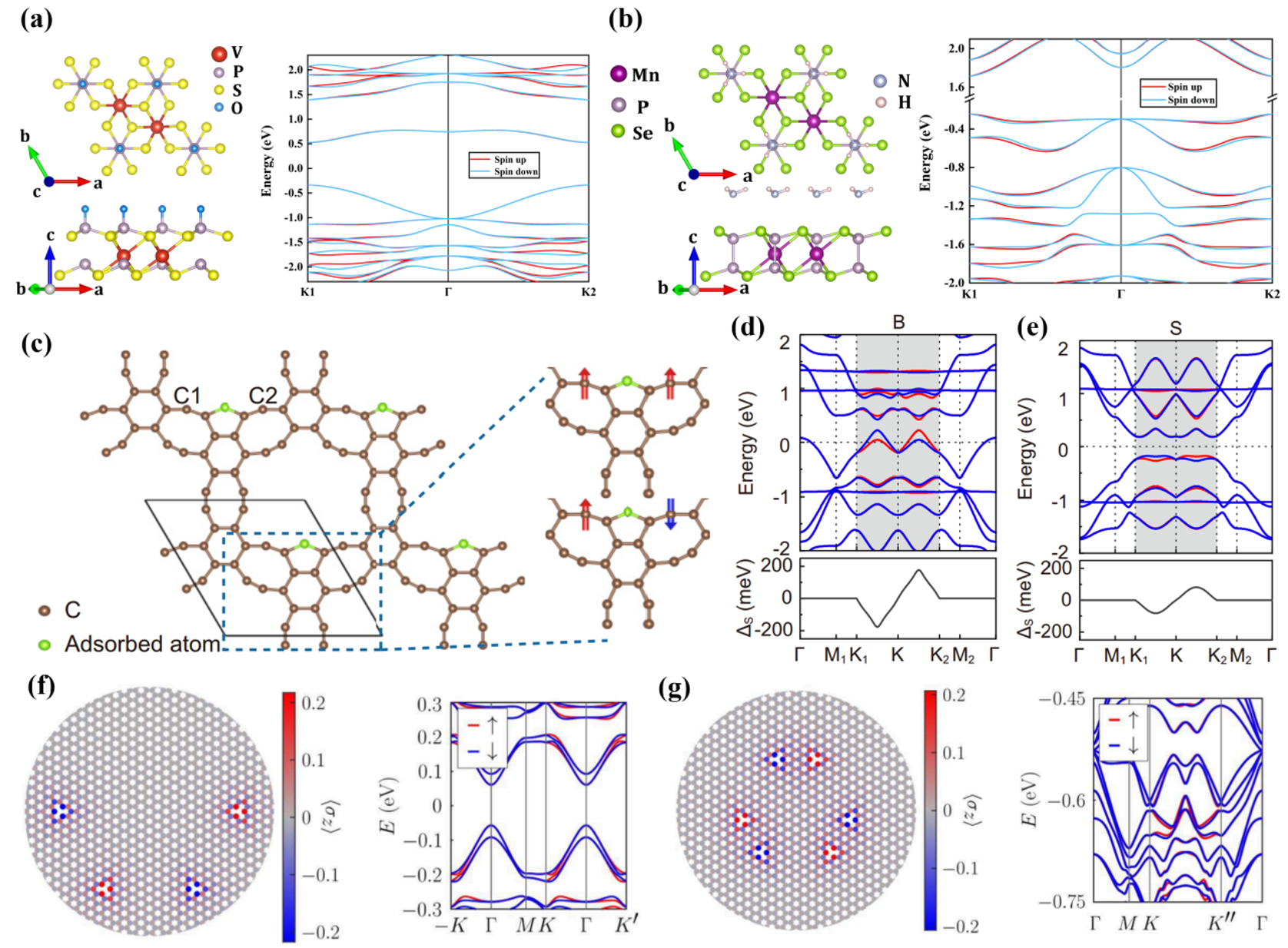}
	\caption{\label{fig:figure5}(a) Structural configuration and corresponding band structure of a monolayer VPS$_3$ with an adsorbed oxygen atom\cite{liu2025unlocking}. (b) Structural configuration and corresponding band structure of a monolayer MnPSe$_3$ with an adsorbed NH$_3$ molecule\cite{liu2025unlocking}. (c) Schematic illustration of two-dimensional hydrogenated graphyne (HGY) along with its ferromagnetic and antiferromagnetic magnetic configurations\cite{72r8-lq5r}. (d) and (e) Band structures of HGY after adsorption of B and S atoms\cite{72r8-lq5r}. (f)-(g) Atomic magnetization and energy bands corresponding to a distinct spin configuration with spin groups $^2$m and $^1$3$^2$m\cite{vina2025building}.}
\end{figure*}
Surface adsorption serves as a versatile strategy to engineer altermagnetism in 2D systems.
Two-dimensional antiferromagnetic materials can host altermagnetism through surface adsorption\cite{liu2025unlocking}. Based on symmetry analysis, altermagnetism can be achieved in a 2D antiferromagnets by adsorbing atoms or molecules at specific sites to break the symmetry that protects spin degeneracy in the band structure, while preserving the rotational or mirror symmetry connecting opposite-spin sublattices. As shown in Figs. \ref{fig:figure5}(a) and (b), taking N\'eel-type antiferromagnets VPS$_3$ and MnPSe$_3$ with spin point group $^2\overline{3}$$^2m$ as examples, O atoms and NH$_3$ molecules were tested at three possible adsorption sites. Calculations of adsorption energy and chemical potential phase diagrams indicate that the highest-coverage adsorption is the most stable. The lowest-energy adsorption site lies exactly on a mirror plane, which breaks the $\mathcal{PT}$ symmetry protecting the spin degeneracy. Since the mirror symmetry connecting the opposite-spin sublattices remains unbroken, a momentum-dependent alternating spin polarization emerges in the absence of net magnetization, and the spin texture exhibits \emph{i}-wave symmetry.

2D hydrognated graphyne (HGY) consists of alternating hexagonal and octagonal carbon rings. Its unique nonlinear \emph{sp} bonds and $\pi$-conjugated framework allow a smooth transition from \emph{sp} to \emph{sp}$^2$ hybridization upon adsorption of nonmagnetic \emph{sp} impurity atoms (e.g., B, N, P, S). When adsorbed at specific sites in the HGY unit cell---namely, at the center bridging two adjacent octagonal rings and directly above a hexagonal ring---these impurity atoms can induce local magnetic moments\cite{72r8-lq5r}, as illustrated in Fig. \ref{fig:figure5}(c). By bridging the octagonal rings and triggering electronic redistribution, they effectively induce collinear compensated spin order with zero net magnetization. Theoretical calculations reveal that B, N, and P atoms convert graphyne from a semiconductor to a metal, while the S-adsorbed system remains semiconducting due to the even number of electrons contributed by S. Moreover, the spin splitting is observed after the adsorption of B and S atoms\cite{72r8-lq5r}. This momentum-dependent spin splitting has magnitudes of 177.4 meV for B and 82.3 meV for S, as shown in Figs. \ref{fig:figure5}(d) and (e). In contrast, N and P atoms fail to induce altermagnetism due to next-nearest-neighbor ferromagnetic coupling.

Moreover, recent work has demonstrated that the adsorption of hydrogen atoms on graphene can induce magnetic moments\cite{vina2025building}. Controlling the number of adsorbed H atoms and symmetry tuning enable the realization of multiple magnetic phases, including  ferromagnetic, antiferromagnetic, ferrimagnetic, and altermagnetic states. Specifically, the covalent bonding between the hydrogen $s$ orbital and the carbon $p_z$ orbital (out-of-plane) effectively removes an electron from the corresponding graphene sublattice, leaving behind an unpaired electron that generates a magnetic moment localized on the complementary sublattice.
To achieve the altermagnetic phase, it is necessary to adsorb at least four H atoms. This ensures the removal of $\mathcal{PT}$ symmetry. Two spin groups are compatible with hydrogenated graphene: one is $^2$m, which has a mirror plane perpendicular to the lattice paired with spin inversion; the other is $^1$3$^2$m, which includes a threefold rotation not paired with spin inversion, and three vertical mirror symmetries that are paired with spin inversion. As illustrated in Figs. \ref{fig:figure5}(f) and (g), the first spin group requires a minimum of four H atoms to realize the altermagnetic phase, while the second requires at least six. Scanning tunneling microscopy (STM) manipulation of individual H atoms has already demonstrated this concept, but the technique cannot yet deliver the periodic arrays required for device-scale films.

\section{\label{sec:level7}Altermagnetism induced by electric field}
\begin{figure*}
	
	\centering
	\includegraphics[width=\linewidth]{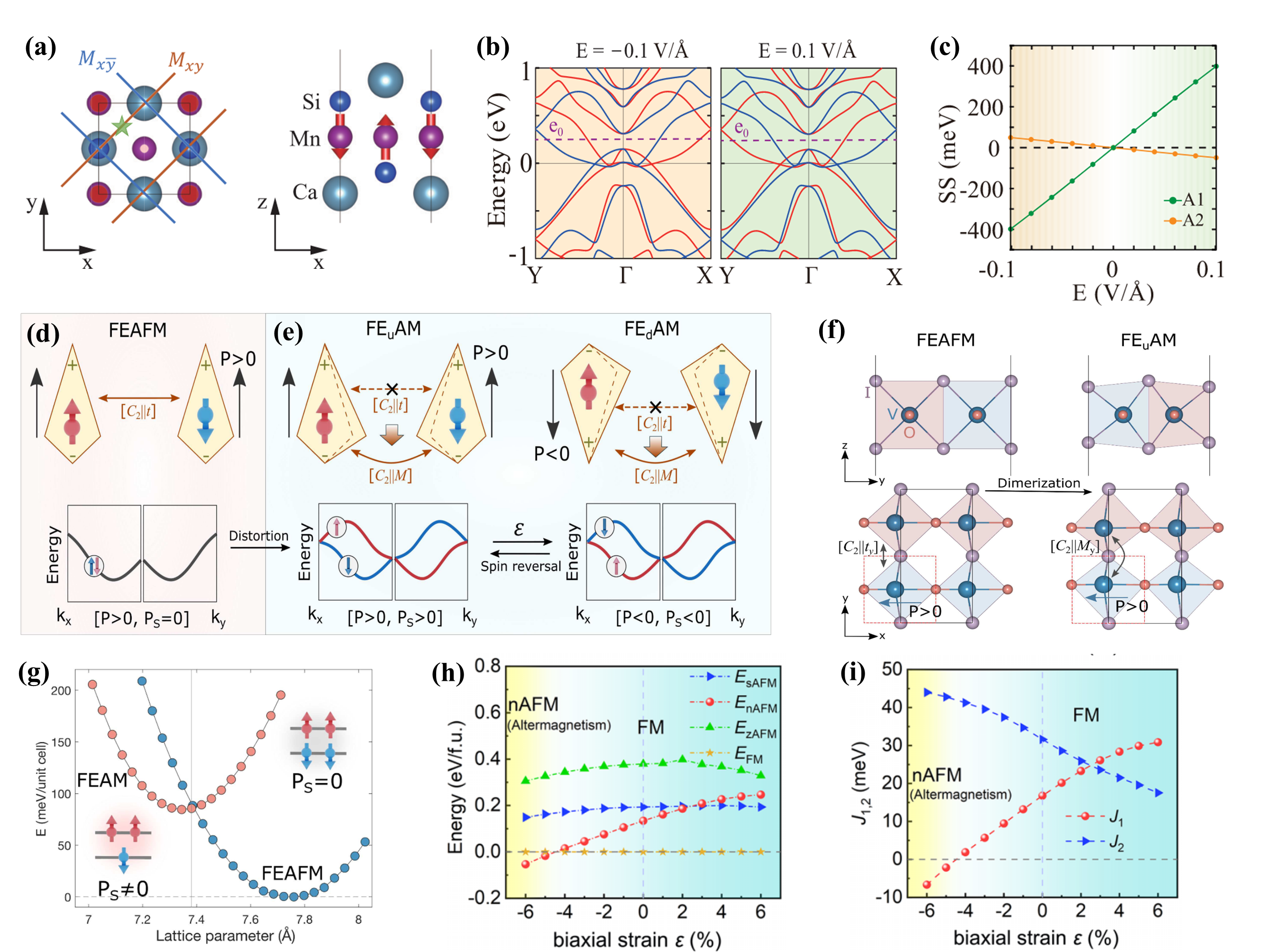}
	\caption{\label{fig:figure6}(a) The crystal structure of CaMnSi\cite{wang2024electric}. (b) The spin-resolved band structure under opposite out-of-plane electric fields\cite{wang2024electric}. (c) the magnitude of spin splitting varies linearly with the strength of electric field\cite{wang2024electric}. (d) and (e) A schematic illustration of structure-distortion-induced altermagnetism\cite{zhu2025two}. (d) Ferroelectric antiferromagnets (FEAFMs) exhibit spin-degenerate band structures protected by translational symmetry. (e) Upon structural distortion, this translational symmetry is broken and replaced by mirror symmetry that links opposite spin sublattices, leading to the emergence of ferroelectric altermagnet ($\mathrm{FE}_{\mathrm{u}}\mathrm{AM}$). When an external electric field is applied, the spin splitting reverses concurrently with the polarization reversal. (f) The crystal structure of VOI$_2$ before and after distortion\cite{zhu2025two}. (g) Energies of monolayer VOI$_2$ in both undistorted and distorted configurations as a function of lattice constant, showing that strain can drive a phase transition between the FEAFM phase and the ferroelectric altermagnet (FEAM) phase\cite{zhu2025two}. (h) Total energies of the four magnetic configurations in the Mn$_2$Se$_2$O as a function of biaxial strain, evaluated relative to the energy of the FM phase. (i) Exchange interaction parameters of the Mn$_2$Se$_2$O as a function of biaxial strain\cite{huangEmergingTwodimensionalHalfmetal2024}.}
	
\end{figure*}
An out-of-plane electric field, a widely used and efficient means to modulate physical properties in 2D materials\cite{wang2018electric}, can break $\mathcal{PT}$ symmetry in 2D antiferromagnets, inducing reversible and tunable spin splitting characteristic. When an out-of-plane electric field is applied, the top and bottom surfaces of the two-dimensional magnet acquire different electrostatic potentials, thereby breaking $\mathcal{PT}$ symmetry. Consequently, in $\mathcal{PT}$-symmetric antiferromagnets whose bands are spin-degenerate, a perpendicular electric field can lift the degeneracy; if the lattice also retains an in-plane rotation or mirror axis, the same field converts the material into an altermagnet. For example, CaMnSi is a $\mathcal{PT}$ antiferromagnet that possesses the $[C_2\,\|\,M_{xy}]$ symmetry, as illustrated in Fig. \ref{fig:figure6}(a). Applying an out-of-plane electric field induces momentum-dependent spin splitting in the band structure, a defining feature of altermagnetism, as illustrated in Fig. \ref{fig:figure6}(b)\cite{wang2024electric}. This occurs because the field breaks $\mathcal{PT}$ symmetry while preserving the $[C_2\,\|\,M_{xy}]$ symmetry. 

Moreover, as illustrated in Fig. \ref{fig:figure6}(b), reversing the field direction simultaneously reverses the nonrelativistic spin splitting, as the two states are connected by the $\mathcal{PT}$ operation, i.e., $\mathcal{PT}\varepsilon_{+E}(s,\textbf{k})=\varepsilon_{-E}(-s,\textbf{k})$. Furthermore, the magnitude of spin splitting varies linearly with the strength of electric field, as shown in Fig. \ref{fig:figure6}(c). By screening two-dimensional material databases, 24 antiferromagnets are predicted to exhibit altermagnetism under an out-of-plane electric field\cite{wang2024electric}. In addition, MnPSe$_3$\cite{mazin2023induced} and VPSe$_3$\cite{shvd-vmvs} are also proposed to realize altermagnetism via this mechanism.

\section{\label{sec:level8}Altermagnetism induced by structure distortion}
Structural distortions in 2D materials can break translational symmetry while preserving ferroelectric order, providing a route to realize ferroelectric altermagnetism\cite{zhu2025two,wang2025two}. It always can arise from Jahn-Teller effects, Peierls-type dimerization, or structural modifications externally tuned by temperature, pressure, or chemical doping\cite{liPhaseTransitions2D2021}. In ferroelectric systems, inversion symmetry is intrinsically broken. However, the magnetic sublattices are typically connected by translational symmetry, giving rise to conventional antiferromagnets, as shown in Fig. \ref{fig:figure6}(d)\cite{zhu2025two}. To realize ferroelectric altermagnets, translational symmetry must be broken while preserving ferroelectric polarization, accompanied by the emergence of a symmetry operation that connects the magnetic sublattices. This can be accomplished through lattice distortions, as illustrated in Fig. \ref{fig:figure6}(e). When these distortions occur perpendicular to the polarization, they break the translational symmetry while preserving the FE order, and emerge mirror symmetry that connects the magnetic sublattices, thereby giving rise to ferroelectric altermagnetism. In this case, the spin splitting of the altermagnet is coupled to the direction of the ferroelectric polarization. Under external fields, the spin splitting reverses concurrently with the ferroelectric polarization reversal. Fig. \ref{fig:figure6}(f) illustrates an example in VOI$_2$, where V-V dimerization breaks translational symmetry while preserving mirror symmetry, thereby driving a transition from the antiferromagnetism to the altermagnetism\cite{zhu2025two}. This approach provides a platform for exploring the interplay between ferroelectricity and altermagnetism. 

\section{\label{sec:level9}Altermagnetism induced by strain}
Strain can induce structure distortions and modulate magnetic exchange interactions, enabling the conversion of ferromagnets or spin-degenerate antiferromagnets into altermagnets. Because 2D materials exhibit exceptional flexibility and strong strain sensitivity, strain engineering emerges as an effective and promising strategy for tuning their electrical, magnetic, and optical properties\cite{qiRecentProgressStrain2023}. Meanwhile, the experimental feasibility of applying controllable strain to 2D materials is well established\cite{yangStrainEngineeringTwodimensional2021}, for example through substrate-induced lattice mismatch. Therefore, it is feasible to generate and modulate altermagnetism in 2D materials via strain. 
	
Recently, there has been growing interest in strain-tuning the properties of 2D altermagnets, including studies on the multipiezo effect\cite{chen2024strain,zhangMultipleStraininducedEffects2025,xunStackingStrainengineeringInduced2025a,liStraininducedValleyPolarization2024}---which encompasses piezovalley responses, piezomagnetism, and piezoelectricity---as well as on strain-induced modulation of their electronic structures\cite{zhangStraintunableSpinvalleyLocking2025}. Furthermore, studies have shown that strain can induce topological phase transitions and give rise to a rich variety of topological states\cite{xunStackingStrainengineeringInduced2025a,fuMultipleTopologicalPhases2025,liStraininducedValleyPolarization2024}. In addition, a strain-induced transition from \emph{g}-wave to \emph{d}-wave altermagnetism has also been proposed\cite{liAltermagnetismStrainInduced2025}.

Strain not only can tune the properties of existing 2D altermagnets, but also serves as an effective mechanism for inducing altermagnetism. From a mechanistic perspective, strain can induce altermagnetism through lattice distortions\cite{zhu2025two} and the modulation of exchange interactions\cite{huangEmergingTwodimensionalHalfmetal2024}. As discussed in the previous section, lattice distortions can break specific symmetries, thereby giving rise to altermagnetism. Such lattice distortions can be generated by applying strain. We next take VOI$_2$ as an illustrative example. As illustrated in Fig. \ref{fig:figure6}(f), VOI$_2$ exhibits two structural phases, whose stability is strongly strain dependent. As shown in Fig. \ref{fig:figure6}(g), tensile strain expands the V-V distances and drives the system into the antiferromagnetism, whereas compressive strain stabilizes the altermagnetism. In this way, strain effectively controls the structural and magnetic landscape of VOI$_2$, enabling reversible switching between the antiferromagnetism and altermagnetism.

Lattice distortions can break the relevant symmetries and convert a spin-degenerate antiferromagnet into an altermagnet, while the modulation of exchange interactions can drive a transition from a ferromagnetic state to an altermagnetic one. Notably, strain-driven transitions between ferromagnetic and antiferromagnetic phases have been widely investigated in earlier works\cite{wangStrainInducedIsostructuralMagnetic2016,cenkerReversibleStraininducedMagnetic2022,calloriStraininducedMagneticPhase2015}. Recent studies have shown that the ferromagnet Mn$_2$Se$_2$O undergoes a transition to an altermagnetic state when applying a compressive strain of about 5\%\cite{huangEmergingTwodimensionalHalfmetal2024}. As shown in Fig. \ref{fig:figure6}(h), compressive strain progressively reduces the energy difference between the ferromagnetic and altermagnetic states, and once the strain exceeds 5\%, the altermagnetic state becomes energetically favored. From the perspective of exchange interactions, Fig. \ref{fig:figure6}(i) reveals that the nearest-neighbor coupling J$_1$ changes from positive to negative under compression, which promotes the stabilization of the altermagnetic state. 
\section{\label{sec:level10}two-dimensional organic altermagnet}
Recently, there has been increasing interest in 2D organic altermagnets\cite{che2025inverse,che2024realizing,ortiz2025organic,v38b-5by1,che2025bilayer}. Based on H\"uckel theory and nonbonding molecular orbitals (NBMOs), a reverse design strategy has been developed, in which organic ligands containing NBMOs are selected and assembled with transition-metal to construct a series of 2D MOF monolayers with altermagnetism\cite{che2025inverse}. In addition, altermagnetism has also been explored in $\pi$-conjugated nanographenes. Nonalternant nanographene units, such as dibenzo[ef,kl]heptalene (DBH), can serve as magnetic building blocks. When assembled into 2D lattices, a $90^\circ$ rotation of the DBH units induces the characteristic momentum-dependent spin splitting of altermagnets\cite{ortiz2025organic}.

\section{\label{sec:level11}perspective}
In this section, we provide a forward-looking perspective on two-dimensional altermagnets, highlighting potential future milestones, promising applications across diverse fields, and possible directions for future research.
\subsection{Experimental: Synthesis and Characterization}
There is no doubt that the field of two-dimensional altermagnets is still in its infancy. Although theoretical advances have been rapid, experimental confirmation remains scarce. The most significant milestone in this field is the first experimental realization of a 2D altermagnet together with definitive evidence of its altermagnetism. Moreover, as outlined in this review, numerous strategies for realizing 2D altermagnetism have been proposed at the theoretical level. Yet compelling experimental evidence supporting their feasibility and validity is still lacking. Demonstrating these strategies experimentally would constitute another major milestone, as it would indicate that 2D altermagnetism can be engineered directly from conventional two-dimensional magnets. Such progress would substantially expand the materials space of 2D altermagnets, providing not only a broad platform for exploring their interplay with other physical properties, but also an enriched reservoir of candidate materials for designing multifunctional devices. From an application perspective, it is also crucial to find air-stable 2D altermagnets with N\'eel temperatures at or above room temperature. Equally important is the ability to produce high-quality atomic layers over large areas. All these developments, if realized, will help propel this emerging field toward its next stage of growth.

Owing to its sensitivity to electron momentum, angle-resolved photoemission spectroscopy (ARPES) enables direct measurements of the electronic structure of crystalline materials. Previous studies have established ARPES as the most direct and powerful technique for identifying altermagnetism\cite{reimers2024direct,krempaskyAltermagneticLiftingKramers2024a,yang2025three,liTopologicalWeylAltermagnetism2025,jiang2025metallic,zhang2025crystal}. One of the key characteristics of altermagnets is the nonrelativistic spin splitting of electronic bands. Consequently, spin-resolved ARPES, which can directly distinguish between spin-up and spin-down states in momentum space, has become the most widely used experimental tool for unambiguously demonstrating altermagnetism\cite{krempaskyAltermagneticLiftingKramers2024a,yang2025three,liTopologicalWeylAltermagnetism2025,jiang2025metallic,zhang2025crystal}. ARPES has also been extensively employed in the study of 2D systems\cite{moAngleresolvedPhotoemissionSpectroscopy2017,cattelanPerspectiveApplicationSpatially2018}. However, whether ARPES techniques, including soft X-ray ARPES and spin-resolved ARPES, can be effectively exploited to provide unambiguous experimental evidence for altermagnetism in 2D materials remains an open and largely unexplored question. In 2D altermagnets, both the magneto-optical Kerr effect\cite{liuUncompensatedLinearDichroism} and anomalous transport properties\cite{PhysRevB.111.184437} have been theoretically predicted. Although these responses do not constitute sufficient proof of altermagnetism, they can serve as indirect or initial experimental evidence. 

\subsection{Spintronics}
Altermagnets hold significant promise for applications in spintronics. Conventional ferromagnet-based spintronic devices face inherent limitations in storage density, operation speed, and robustness. Meanwhile, conventional antiferromagnets, with spin-degenerate electronic structures, are limited in their ability to generate large spin currents. As a result, altermagnets, integrating key advantages of both ferromagnetic and antiferromagnetic systems, are anticipated to be highly promising for the future spintronic applications. A prototypical spintronic device is the spin valve, comprising two magnetic conducting layers separated by a nonmagnetic spacer. Depending on whether the spacer is metallic or insulating, the device exhibits giant magnetoresistance (GMR) or tunneling magnetoresistance (TMR), respectively. Recent studies indicate that, in contrast to conventional magnetic tunnel junctions (MTJs), altermagnetic tunnel junctions (AMTJs) exhibit an anomalous scaling behavior, where TMR diminishes as the barrier thickness increases\cite{yangUnconventionalThicknessScaling2025}. Consequently, the TMR is expected to achieve its largest value in the monolayer limit of the barrier. This is especially beneficial in 2D materials, which allow precise layer control and nearly perfect interfaces in van der Waals heterostructures. Taken together, these features suggest that AMTJs based on 2D altermagnets can achieve superior performance compared to those based on 3D altermagnets. Further experimental confirmation of this anomalous scaling behavior would be highly valuable, as it could contribute to guiding the design of AMTJs based on 2D altermagnets. In addition, large TMR has been theoretically predicted in various AMTJs based on 2D altermagnets\cite{wang2025pentagonal,yangUnconventionalThicknessScaling2025}, which call for experimental validation. A natural next step for research is to realize these AMTJs as building blocks in magnetic random access memories (MRAM) devices, unlocking their potential for high-performance, robust, and high-density spintronic memory applications. Without a doubt, realizing this would constitute a significant milestone in the application of 2D altermagnets.

Recent pioneering work has identified spin-split floating edge states in two-dimensional altermagnetic second-order topological insulators\cite{fangEdgetronicsTwoDimensionalAltermagnets2025}. Exploiting these states has led to the proposal of an unconventional edge tunnel magnetoresistance effect, which offers an electrically switchable platform with an ultrahigh on/off ratio for next-generation spintronic devices\cite{fangEdgetronicsTwoDimensionalAltermagnets2025}. In addition, spin-selective quantum transport in heterostructures composed of a normal metal and a two-dimensional \emph{d}-wave altermagnet has been theoretically investigated, revealing a Fermi-surface-geometry-driven mechanism for achieving perfect spin polarization\cite{fuAllelectricallyControlledSpintronics2025a}. Building on this mechanism, all-electrically controlled spin filters and spin valves have been proposed, establishing a promising platform for scalable, magnetic-field-free spintronic devices\cite{fuAllelectricallyControlledSpintronics2025a}. These results further highlight the advantages of 2D altermagnets for spintronic applications and provide valuable insights for the design of next-generation spintronic devices.
	
The successful realization of 2D spintronic devices and circuits requires the ability to achieve spin-charge conversion, spin transport, and spin manipulation\cite{linTwodimensionalSpintronicsLowpower2019}. However, current studies on 2D altermagnets have predominantly focused on spin transport, while comparatively little attention has been paid to the realization of spin-charge conversion and spin manipulation, as well as to the corresponding device concepts. Therefore, corresponding studies deserve increased attention in future research. Another promising direction that deserves attention is magnetic skyrmions. Because of the vanishing topological charge, the skyrmion Hall effect has long been considered absent in antiferromagnets. However, recent studies predict that this effect can emerge in 2D altermagnets\cite{jinSkyrmionHallEffect2024}. This intriguing prediction awaits experimental validation. As the most prominent application of skyrmions is racetrack memory\cite{back2020SkyrmionicsRoadmap2020}, developing racetrack devices based on 2D altermagnetic skyrmions should become a key focus of future research. Whether other two-dimensional magnetic topological quasiparticles can be realized in 2D altermagnets remains an open question. It is also worth exploring whether phenomena previously realized in 2D magnetic systems can be extended to 2D altermagnets. For example, four distinct resistance states have been achieved in the tunnel junctions composed of Fe$_n$GeTe$_2$ layers separated by a 2D ferroelectric In$_2$Se$_3$ barrier\cite{suVanWaalsMultiferroic2021}, raising the question of whether analogous functionalities can be realized in altermagnetic counterparts.

\subsection{Topology, Cornertronics and Superconductivity}
Substantial progress has been made in understanding the interplay between magnetism and nontrivial band topology\cite{wangIntrinsicMagneticTopological2023, liuMagneticTopologicalInsulator2023,bernevigProgressProspectsMagnetic2022}, providing an established context for discussing topological responses in 2D altermagnets. To date, a growing number of two-dimensional altermagnetic topological materials have been proposed, including, but not limited to, CrO\cite{guo2023quantum,chen2023giant}, V$_2$Te$_2$O\cite{PhysRevB.111.184437,PhysRevB.108.024410}, MnS$_2$\cite{wang2025pentagonal}, and Cr$_2$BAl\cite{sattigeri2025dirac}. The quantum anomalous Hall effect (QAHE) is a topological quantum transport phenomenon in the absence of an external magnetic field, and it holds significant promise for low-dissipation electronics and topological quantum devices. Recently, the QAHE has been theoretically predicted in 2D altermagnets\cite{guo2023quantum,wuQuantumAnomalousHall2023}, representing the first realization of this effect in collinear antiferromagnetic systems. These predictions call for experimental efforts to realize the QAHE in 2D altermagnets, which would not only confirm the theoretical proposals but also open new avenues for antiferromagnetic topological electronics. In addition, the symmetry requirements for topologically protected spin-polarized Dirac points in 2D altermagnets have been systematically analyzed, and the corresponding altermagnetic spin-wallpaper groups have been identified\cite{parshukovTopologicalCrossingsTwodimensional2025}. In the future, high-throughput first-principles calculations could be employed to systematically search for 2D altermagnetic materials with specific topological properties. It remains an open question whether stacking configurations and external electric fields can modulate the topological phases of 2D altermagnets.

Cornertronics is an emerging research field that exploits symmetry-protected electronic states localized at the corners of higher-order topological insulators, known as corner states, as controllable degrees of freedom for emerging electronic functionalities. In this context, 2D altermagnets have recently demonstrated unique potential in this direction. On the one hand, a spin-corner locking mechanism has been identified in second-order topological insulators (SOTIs) realized in 2D altermagnetic systems\cite{liNearlyFlatEdges2025,yangSpinSelectiveSecondOrderTopological2025,wang2025pentagonal}. As a result, 2D altermagnets constitute an ideal platform for realizing spin-mediated higher-order topological corner states. Meanwhile, intrinsic spin-corner coupling is regarded as a characteristic of 2D altermagnetic SOTIs. In bilayer NiZrI$_6$ nanodisks, a spin-corner-layer locking mechanism is realized by introducing the layer degree of freedom\cite{wangElectricFieldControlTerahertz2025}. On the other hand, corner states can be flexibly manipulated by external fields, such as electric fields and strain. For example, in bilayer NiZrI$_6$ nanodisks, a vertical electric field enables simultaneous reversal of the spin and layer polarizations of corner states, thereby modulating the absorption, emission, and polarization  of terahertz waves\cite{wangElectricFieldControlTerahertz2025}. Furthermore, in 2D altermagnets CrO and Cr$_2$Se$_2$O, uniaxial strain breaks the M$_{xy}$ symmetry, leading to spin-resolved corner modes and a corner-polarized second-order topological insulator\cite{yangSpinSelectiveSecondOrderTopological2025}. These results suggest that 2D altermagnets provide a fertile playground for cornertronics. Looking ahead, the rich interplay between topology and altermagnetism is expected to further expand the scope of cornertronics in two-dimensional systems.

The interplay between altermagnetism and superconductivity has attracted growing research interest in recent years, owing to the intrinsic spin-momentum locking characteristic of altermagnets\cite{liuAltermagnetismSuperconductivityShort2025}. In this context, a variety of intriguing phenomena have been theoretically predicted, including the superconducting diode effect\cite{PhysRevB.110.024503,cv8s-tk4c}, Josephson effects\cite{ouassouDcJosephsonEffect2023a}, Andreev reflection\cite{sunAndreevReflectionAltermagnets2023,PhysRevB.108.L060508}, and finite-momentum superconductivity\cite{mukasaFinitemomentumSuperconductivityTwodimensional2025,hongUnconventional,zhangFinitemomentumCooperPairing2024,chakrabortyZerofieldFinitemomentumFieldinduced2024}. Meanwhile, altermagnets provide a fertile platform for exploring unconventional and topological superconductivity. A strong exchange field can drive the emergence of Bogoliubov Fermi surfaces in altermagnets, which suppress conventional spin-singlet \emph{s}-wave pairing and ultimately induce chiral \emph{p}-wave superconductivity or finite-momentum superconductivity\cite{hongUnconventional}. In 2D metals with \emph{d}-wave altermagnetism and Rashba spin-orbit coupling, the system favors mixed spin-singlet \emph{s}-wave and spin-triplet \emph{p}-wave pairings, and when \emph{p}-wave pairing dominates, it can host various topological superconducting phases\cite{PhysRevB.108.184505}. Recent studies on altermagnet-superconductor interfaces and heterostructures have revealed a variety of nontrivial superconducting interface effects, including orientation-dependent Andreev reflection\cite{sunAndreevReflectionAltermagnets2023}, finite-momentum Cooper pairing\cite{zhangFinitemomentumCooperPairing2024}, gapless superconducting states\cite{weiGaplessSuperconductingState2024,weiUnconventionalHallEffect2025}, as well as higher-order topological superconductivity supporting Majorana corner modes\cite{liMajoranaCornerModes2023}. In the two-dimensional limit, their atomically sharp interfaces and flexible van der Waals stacking markedly enhance proximity coupling, positioning 2D altermagnets as a promising platform for superconducting spintronic devices.

\subsection{Valleytronics and multiferroics}
Valleytronics is a field of research that exploits the valley degree of freedom in momentum space as an information carrier, with the goal of enabling low-power and high-efficiency information encoding, processing, and storage. 2D materials are regarded as an ideal platform for valleytronics\cite{xuValleytronicsFundamentalChallenges2025}, which provides a solid basis for exploring valley-dependent phenomena in 2D altermagnets and their potential valleytronic applications. In 2D altermagnets, spin-momentum locking protected by spin-group symmetries naturally gives rise to a theoretical basis for crystal-symmetry-protected spin-valley locking, also referred to as C-paired spin-valley locking\cite{ma2021multifunctional}. Since the spin-valley locking is protected by crystal symmetries, various external approaches, such as strain and electric fields, can be employed to manipulate the valley degree of freedom, thereby giving rise to a variety of intriguing physical phenomena. For example, valley polarization can be realized in 2D altermagnets by applying strain\cite{PhysRevB.110.184408,zhangStraintunableSpinvalleyLocking2025} or external electric fields\cite{PhysRevB.110.L220402}. Meanwhile, the crystal valley Hall effect has been proposed and is predicted to be realized in 2D altermagnets Fe$_2$WSe$_4$ and Fe$_2$WS$_4$\cite{PhysRevB.111.094411}. These phenomena call for further experimental investigation. Conventional valleytronic materials have long faced challenges including limited valley lifetimes, insufficient material quality and scalability, and low operating temperatures, which collectively hinder the practical implementation of valleytronic applications\cite{xuValleytronicsFundamentalChallenges2025}. Whether 2D valleytronic altermagnets suffer from similar limitations, or instead offer viable routes to overcome these challenges, remains an open question that warrants further investigation.

Electrical control of magnetism provides an energy-efficient route for tuning magnetism-related physical properties, because the absence of heating by currents. In multiferroic materials, this control can be realized through magnetoelectric coupling mechanisms. At present, theoretically proposed 2D multiferroic altermagnets include ferroelectric altermagnets\cite{wang2025two,zhu2025two}, in which ferroelectricity coexists with altermagnetism, and antiferroelectric altermagnets\cite{PhysRevLett.134.106801}, where antiferroelectric order is combined with altermagnetism. Among these systems, the ferroelectric altermagnets, which conmbine  altermagnetism and sliding ferroelectricity, is particularly promising for applications, owing to its enhanced fatigue resistance and low energy consumption. Moreover, an unconventional magnetoelectric coupling mechanism has been proposed, in which the reversal of nonrelativistic spin splitting is accompanied by ferroelectric switching\cite{zhu2025sliding,c4dm-12ng,sun2024altermagnetism}. A detailed discussion of these can be found in Section \ref{sec:level4}. Experimental validation of these results would be a crucial step toward enabling practical applications of multiferroic altermagnets. In addition, electrostatic doping offers an alternative route to achieving electrical control of magnetism\cite{wang2022magnetic}. However, this approach remains largely unexplored in two-dimensional altermagnets.

\subsection{Miscellaneous}
Future research should not only aim to expand the library of candidate materials, but also to deepen our understanding of their underlying physics. Developing symmetry-guided effective theories---such as $k\cdot p$ models for \emph{d}-wave altermagnets\cite{wangSymmetryConstrainedAnomalousTransport2025}---and benchmarking them against first-principles calculations will be essential for uncovering the microscopic origins of intrinsic transport phenomena and for guiding experimental probes. Moreover, the design of novel spintronic devices based on 2D altermagnets remains to be explored. Furthermore, integrating 2D altermagnets with other research areas could yield a wider variety of emergent physical phenomena and represents a promising direction for future studies. In addition, it remains an open question whether recently proposed symmetry-driven unconventional magnetism\cite{liu2025different}, such as 2D fully compensated ferrimagnets\cite{PhysRevLett.134.116703}, Type IV 2D Collinear Magnets\cite{rn1l-d6cq} and \emph{p}-wave magnets\cite{hellenes2023p}, can be realized using the design strategies summarized in this review. Continued exploration of 2D altermagnetism promises to unveil new physics and to spawn spintronic devices whose efficiency and functionality exceed current limits, positioning 2D altermagnets at the forefront of future magnetism research and quantum technologies. 
\section{\label{sec:level12}conclusion and challenges}
This review centers on the symmetry principles and material realizations of two-dimensional altermagnets. It highlights their symmetry descriptions and classifications, compiles the theoretically predicted candidates reported so far, and discusses diverse design strategies for achieving altermagnetism in 2D materials. Finally, we offer a perspective for the field and highlight several promising directions where future breakthroughs may emerge. Again, the field is still in its infancy: most predictions rest on calculations, and the key experiments that would confirm---or refute---two-dimensional altermagnetism in the proposed candidates have yet to be performed. Therefore, advancing experimental efforts will be crucial for future progress. 

Research on 2D altermagnets has already made notable progress, and the rapidly growing body of theoretical and computational studies reflects the surging interest and momentum in this emerging area. Nevertheless, many challenges still remain in achieving a deeper understanding of 2D altermagnetism and in predicting new 2D altermagnets. In magnetic systems, particularly those containing transition-metal or rare-earth elements, incorporating a Hubbard U correction is often essential in first-principles calculations. However, the choice of U value can strongly influence the calculated electronic and magnetic properties. Key magnetic quantities, such as exchange splitting, local magnetic moments, interlayer exchange interactions, and even the relative energetic stability of different magnetic configurations, can exhibit pronounced U-dependence. For 2D altermagnets, where the prediction fundamentally relies on an unambiguous identification of the magnetic ground state, this U sensitivity introduces non-negligible uncertainty. Moreover, the choice of the DFT+U scheme, whether the Dudarev's or the Liechtenstein's approach, can also affect the correct prediction of the magnetic ground state, as demonstrated in the case of monolayer CrOCl\cite{zhuMagneticGroundState2023}.

At the same time, a broadly applicable strategy for predicting the true magnetic ground state is still lacking. As summarized in Section \ref{sec:level3}, most predicted altermagnets have had their ground states determined by comparing the energies of only a few collinear spin configurations. However, the magnetic phase space is inherently large, encompassing collinear, coplanar, and noncoplanar arrangements. If the sampled configurations fail to include the actual ground state, the prediction is inevitably incorrect. Although spin-spiral calculations have shown promise in identifying magnetic ground states in two-dimensional magnets\cite{sodequist2024two,sodequist2024magnetic}, this approach is limited by its requirement of a primitive cell containing only one magnetic atom, which limits its applicability to more complex materials. 

In most theoretical studies of two-dimensional altermagnets, spin-orbit coupling (SOC) is typically neglected. However, this approximation can lead to substantial inaccuracies in certain cases. For example, it has been reported that materials predicted to be ferromagnetic in the absence of SOC actually develop a chiral spin-spiral ground state once SOC is taken into account\cite{sodequist2024magnetic}. As discussed in Section \ref{sec:level4}, a similar concern arises in the twisted 90$^\circ$ bilayer CrSBr, where the easy axes of the two layers are mutually orthogonal. Consequently, whether a collinear spin configuration can remain stable after SOC is included remains an unresolved issue that warrants careful examination. Further studies on altermagnets should take into account whether the effects induced by SOC can be neglected.
\begin{acknowledgments}
This work is financially supported by the National Natural Science Foundation of China (Grant No. 12474229 and 12504268), the Fundamental Research Funds for the Central Universities(No. 2025ZYGXZR039) and the Guangdong Provincial Key Laboratory of  Functional and Intelligent Hybrid Materials and Devices (Grant No. 2023-GDKLFIHMD-04).
\end{acknowledgments}
\bibliography{ref}

\end{document}